\documentstyle[eqsecnum,aps,epsf,prl]{revtex}
\begin{document}
\title{Neutral weak currents in pion electroproduction on the nucleon.}
\author{Michail P. Rekalo \footnote{ Permanent address:
\it National Science Center KFTI, 310108 Kharkov, Ukraine}
}
\address{Middle East Technical University, 
Physics Department, Ankara 06531, Turkey}
\author{Jacques Arvieux}
\address{IPNO, IN2P3, BP 1, 91406 Orsay, France}
\author{Egle Tomasi-Gustafsson}
\address{ DAPNIA/SPhN, CEA/Saclay, 91191 Gif-sur-Yvette Cedex, 
France}
\def\qqbar{$<q\bar{q}>$ }
\def\ssbar{$<s\bar{s}>$ }
\def\q{quark} 
\def\Q2{$Q^2$}
\def\scalar{$< N | \bar{s}s | N >$ }
\def\vector{$< N | \bar s\gamma_{\mu}{s} | N >$ }
\def\axial{$< N | \bar s\gamma_{\mu} \gamma_5 {s} | N >$ }
\def\geg{$G_E^\gamma$ }
\def\gmg{$G_E^\gamma$ }
\def\gez{$G_E^Z$ }
\def\gmz{$G_M^Z$ }
\def\ges{$G_E^s$ }
\def\gms{$G_M^s$ }
\def\gad{$G_{N\Delta}^A$ }
\def\F1g{$F_1^\gamma$ }
\def\F2g{$F_2^\gamma$ }
\def\f1z{$F_1^Z$ }
\def\f2z{$F_2^Z$ }
\def\f1s{$F_1^s$ }
\def\g2s{$F_2^s$ }
\def\G0{$G^0$ }
\def\Z0{$Z^0$ }

\maketitle
\date{\today}
\begin{abstract}
Parity violating asymmetry in inclusive scattering of longitudinally 
polarized 
electrons by unpolarized protons with $\pi^0$ or $\pi^+$ meson production, is 
calculated as a function of the momentum transfer squared $Q^2$ and the total 
energy $W$ of the $\pi N$-system. This asymmetry, which is induced by the 
interference of the one-photon exchange amplitude with the parity-odd  
part of the 
$Z^0$-exchange amplitude, is calculated for the $\gamma^*(Z^*)+p\to N+\pi$ 
processes ($\gamma^*$ is a virtual photon and $Z^*$ a virtual Z-boson)
considering the $\Delta$-contribution in the $s-$channel, the standard 
Born contributions and vector meson ($\rho$ and $\omega$) exchanges in 
the 
$t-$channel. Taking into account the known isotopic properties of the hadron 
electromagnetic and neutral currents, we show that the P-odd term is the sum 
of 
two contributions. The main term is model independent and it can be 
calculated 
exactly 
in terms of fundamental constants. It is found to be linear in $Q^2$. The 
second 
term 
is a relatively small correction which is determined by the 
isoscalar component of the electromagnetic current. Near threshold and in the 
$\Delta$-region, this isoscalar part is much smaller (in 
absolute value) than the isovector one: its contribution to the asymmetry 
depend on 
the polarization state (longitudinal or transverse) of the virtual photon.
\end{abstract}

\section{Introduction}
Parity violation (PV) was discovered in 1956 in nuclear beta-decay by C.S. Wu 
\cite{Wu57}, following a suggestion of T.D. Lee and C. N. Yang \cite{LY56}. In 
1960, Ya. Zeldovich \cite{Ze59} pointed out that PV should lead to parity-odd 
(P-odd) terms also in electron-hadron interactions. These are now considered as 
a 
manifestation of the electroweak interaction, whose properties are dictated 
by 
the Standard Model (SM). Several P-odd observables have  since been studied , 
in 
two types of PV experiments, namely in atomic physics \cite{WCS,BB97} (at 
very 
low energy and momentum transfer) and in electron scattering (at relatively 
high 
energies and non-zero momentum transfers).

At first, these experiments were aiming at testing the SM and measuring the 
Weinberg angle. A pioneering experiment was performed at SLAC on a deuterium 
target \cite{Pr78}, followed 10 years later by experiments at Mainz on $^9$Be 
\cite{He79} and Bates on $^{12}$C \cite{So90}. Their determination of the 
Weinberg angle were confirmed later on, within their stated accuracy of 10\%, 
by 
high energy experiments. Since $sin^2\theta_W$ is now known to three decimal 
places [$sin^2\theta_W = 0.23124(24)$] \cite{Co98}, the emphasis of e-p 
scattering today, is to make  use of the SM to learn about the internal 
structure of the nucleon.

Until recently, it has been assumed that the nucleon was only made of u and d 
valence or sea quarks, but there are indications that the nucleon carries 
also 
hidden strangeness:
\begin{itemize}
\item the sigma-term (deduced from the pion-nucleon scattering length) is 
very different from the theoretical value calculated within the chiral 
perturbation theory (which is a realization of the SM at low energy), 
indicating 
that $35\%$ of the nucleon mass might be carried out by strange quarks 
\cite{Ch76,DN76,GLS91},
\item experiments of polarized Deep-Inelastic-Scattering (DIS) of leptons 
show 
that up to 10-20$\%$ of the nucleon spin could be carried by strange quarks 
\cite{As88,Ad94,Ab95,Ad93},
\item elastic scattering of neutrinos and anti-neutrinos by protons can only 
be explained by taking into account strange quarks in the nucleon 
\cite{Ah87,GLW93},
\item a natural explanation of the strong violation of the OZI-rule in 
$p\overline{p}$ annihilation \cite{El95,RAT97a}  and of $\phi$- production 
\cite{RAT97b} or $\eta$-meson production \cite{RAT97c} in nucleon-nucleon 
interactions takes into account a nucleon (antinucleon) strange sea.
\end{itemize}

These experiments are sensitive to various aspects of nucleon structure: for 
example, the sigma-term and the results of $NN$ or $N\overline{N}$ experiments 
are 
sensitive 
to the scalar part of the hadronic current, polarized DIS and elastic scattering 
of neutrinos (or anti-neutrinos) by protons are sensitive to the 
vector-axial current.
In this respect, PV in electron-nucleon scattering seems the most attractive 
way 
of measuring the {\it strange} vector current, thanks to a clean theoretical 
interpretation through the SM.

The SAMPLE collaboration at MIT-Bates, has measured PV asymmetries in $\vec e p$ 
elastic scattering at 
$Q^2$= 0.1 $(GeV/c)^2$ ($Q^2=|k^2|= -k^2$, where $k$ is the four-momentum 
transfer squared)
and large angle \cite{Sp00}, which allowed them to 
obtain 
the first experimental determination of the weak magnetic form-factor of the 
proton. 
From this measurement and the knowledge of the proton and neutron 
electromagnetic form-factors, one could extract a strange magnetic 
form-factor 
\gms =$(0.61 \pm 0.17 \pm 0.21)~\mu_N$.
Note that most 
calculations based on QCD or quark models predict negative values for 
\gms (see e.g. \cite{Ja89,MB94,We95,Ha96,Ki97,Le96,MI97,MK99,DLW98}).

Another experiment, done by the HAPPEX collaboration at Jefferson 
Lab,\cite{An99} has done a measurement at \Q2= 0.48 $(GeV/c)^2$ and small 
scattering angle $\theta_e= 35^0$ where the sensitivity to the weak electric 
form 
factor \gez is enhanced. Here the measured asymmetry $A = (-14.2 \pm 
2.2)\cdot10^{-6}$ is consistent with the SM prediction in the absence of 
\ssbar 
components in the nucleon sea. From this asymmetry, one can deduce the 
following contribution to the strange form-factor:
$$G^s_{HAPPEX} = G_E^s + 0.39 G_M^s  = (0.023 \pm 0.034 \pm 0.022\pm 
0.026)~\mu_N,$$
compatible with zero within the error bars.

These results have stimulated a strong  
interest and many predictions, for both \ges  and \gms,
have been published, whether within quark models \cite{Ja89}, Chiral 
Perturbation Theories \cite{We95,Ha96}  or Lattice QCD calculations \cite{Le96}. 
These 
calculations predict that while \gms is essentially constant as a function of 
\Q2, \ges may vary rapidly. They also indicate that there might be some 
cancellation between \ges  and \gms which are predicted of different signs. 
Therefore new e-p experiments are being set up in order to  check these 
predictions: at $Q^2 = 0.225$ (GeV/c)$^2$ at Mainz\cite{PVA4}, at $Q^2$ 
= 
0.1 (GeV/c)$^2$  and forward angles by the HAPPEX collaboration in order to do a 
Rosenbluth separation of \ges  and \gms in combination with the SAMPLE 
results, 
and finally a full separation of \ges  and \gms in the momentum transfer 
range 
$Q^2$ = 0.12-1.0 (GeV/c)$^2$ is foreseen by the \G0 collaboration at 
Jefferson 
Lab \cite{G0}.

It should be stressed that, due to their present high level of precision and 
their firm theoretical basis, PV experiments can also be used to answer other 
important physics questions, besides the existence of a possible \ssbar 
component 
in the nucleon. Let's mention for example:

\hspace{1cm}$\bullet$ search and test for new physics beyond the SM 
\cite{Ro98,PT96,MR96,Mu94,Muk98,RMxx,EL99}. Such effects could manifest 
themselves through, e.g. extra $Z'$-bosons (heavier than the standard \Z0) or 
leptoquarks indicating that there are substructures common to both leptons 
and 
quarks. The use of the \G0 experimental set-up for such studies is being 
discussed \cite{Ca99},

\hspace{1cm}$\bullet$ study of the axial part of the hadronic weak neutral 
current: It has been shown that, for the special case of a $I=3/2$, $J=3/2$ 
spin-isospin transition ($\Delta$-excitation), there is a particular 
sensitivity 
to \gad, the axial-vector transition form-factor which could be determined 
independently of PCAC and free of uncertainties from extrapolation of low 
energy 
theorems \cite{WSK97},  

\hspace{1cm}$\bullet$ measurement of the neutron charge form-factor 
\cite{Fe75,Wa77,AM99}.

The reactions $e+p\to e+p+\pi^0$ and $e+p\to e+n+\pi^+$ are of practical 
interest for experimentalists as they may contaminate the elastic peak. It is 
therefore important to determine their own asymmetries since, if they are much 
larger than or, even, of different  sign from the elastic one,  they might be 
a 
source of errors or large uncertainties.

In 3-body reactions, besides the weak PV asymmetries, there are also strong  
(parity-conserving) interactions, due to the so-called $5^{th}$ response 
function  \cite{PV89}, which are generally much larger (of the order of 
$10^{-2}-10^{-3}$ instead of $10^{-5}-10^{-7}$) than PV asymmetries but which 
cancel in inclusive reactions or when detectors have an azimuthal symmetry. 

Pion production has been studied previously 
\cite{Na82,Jo80}  in quasi 2-body models with stable $\Delta$ isobar.
A more complete calculation including background (Born) terms with 
pseudoscalar 
$\pi N$ coupling with the $\Delta$ treated as a Rarita-Schwinger field with 
phenomenological $\pi N$ electromagnetic transition currents can be found in 
\cite{HD95}.

In the present study, we calculate  PV asymmetries in inclusive 
$N ( e, e') N \pi$ electroproduction, starting from threshold up to the 
$\Delta$-region in an approach similar to the one of ref. \cite{HD95}, but 
differing in the following aspects:
\begin{itemize}
\item the main improvement consists in including $\omega$- and 
$\rho$-exchange 
in the $t$-channel for $\gamma^* (Z^*) + N \to\pi + N$  (where $\gamma^* (Z^*)$ 
is 
a virtual photon  or boson),
\item we use a different parametrization for the $\Delta$ contribution, and 
slightly different values of mass and width,
\item crossing symmetries are treated differently (and less accurately) than 
in 
ref \cite{HD95},
\item we use a pseudoscalar $\pi NN$ interaction in order to identify 
possible 
off-mass-shell effects.

Two remarkable results were found in the calculations of ref.\cite{HD95}, for 
which no explanation or discussion was given:
\begin{itemize}
\item 	the full (background + resonance) asymmetry A does not depend on the 
total energy of the hadronic system, although the separate terms show strong 
(and opposite) variations,
\item	$A$ is, to a good approximation, a linear function of \Q2. 
\end{itemize}
\end{itemize}
The present work agrees with these features. Moreover it gives a physical 
explanation to them as 
it shows that there is a specific parametrization of the asymmetry which 
allows 
to separate the main (isovector) contribution in a model independent way. 
This 
contribution only depends on the Fermi constant $G_F$, the fine structure 
constant $\alpha$ and $sin^2\theta_W$. Therefore its dependence on the 
kinematics can be predicted exactly. Small corrections to the main 
contribution, 
due to the isoscalar part of the neutral vector and axial currents can then be 
calculated and their physical importance assessed.

\section{P-odd beam asymmetry for  $\lowercase{e^-}+N \to 
\lowercase{e^-}+N+\pi$ 
}

We shall consider here the processes ${e^-}+N \to {e^-}+N+\pi$, where $N$ is 
a 
nucleon ($p$ or $n$) and $\pi$ is a pion ($\pi^0$ or $\pi^+$). We take into 
account two standard mechanisms, $\gamma-$ and $Z-$ boson exchanges (Fig. 1), 
predicted by the SM. The matrix element can be written in the following form:
$${\cal M}={\cal M}_{\gamma}+{\cal M}_{Z},$$
\begin{equation}
{\cal M}_{\gamma}= -\displaystyle\frac{e^2}{k^2}\ell_{\mu}{\cal 
J}_{\mu}^{(em)},
\label{eq:eq1}
\end{equation}
$${\cal M}_{Z}=\displaystyle\frac{G_F}{2\sqrt{2}}\left ( g_v^{(e)}\ell_{\mu}+
g_a^{(e)}\ell_{\mu,5}\right )\left({\cal J}_{\mu}^{(nc)}+ {\cal 
J}_{\mu,5}^{(nc)}\right ),$$
where $G_F$ is the Fermi constant of the weak interaction, 
${\cal J}_{\mu}^{(em)}$ is the electromagnetic current for 
$\gamma^*+N\to N+\pi$, ${\cal J}_{\mu}^{(nc)}$ and  ${\cal J}_{\mu 
,5}^{(nc)}$ 
are the vector and vector-axial parts of the neutral weak current for $Z^*+N\to 
N+\pi$. The four-vector $\ell_{\mu}$ and $\ell_{\mu ,5}$ are the 
vector 
and vector-axial 
parts of the neutral weak current of a point-like electron:
$$\ell_{\mu}=\overline{u}(k_2)\gamma_{\mu}u(k_1),$$
\begin{equation}
\ell_{\mu,5}=\overline{u}(k_2)\gamma_{5}\gamma_{\mu}u(k_1)
\label{eq:eq2}
\end{equation}
where $k_1~(k_2)$ is the four-momentum of the initial (final) electron. The 
notation for the particle four momenta is explained in Fig. 1. Note that the 
formula for ${\cal M}_Z$, Eq. (\ref{eq:eq1}), is valid in the so-called local 
limit, where 
$$-k^2 \ll M_Z^2 \simeq 8100~ \mbox{(GeV/c)}^2.$$ 
In the Standard Model the constants $g_a^{(e)}$ and $g_v^{(e)}$ are determined 
by the following expressions:
\begin{equation}
g_a^{(e)}=1,~~g_v^{(e)}=1-4\sin^2\theta_W,
\label{eq:eq3}
\end{equation}
where $\theta_W$ is the 
Weinberg angle. Then:
\begin{equation}
g_v^{(e)}\simeq 0.076, \ i.e.\  g_v^{(e)}\ll g_a^{(e)}.
\label{eq:eq4}
\end{equation}
From Eq. (\ref{eq:eq1}) it appears that P-odd effects in $e^-+N\to e^-+N+\pi$ 
result 
from 
the 
interference between ${\cal M}_{\gamma}$ and ${\cal M}_Z^{(-)}$, which is the 
parity-violating part of ${\cal M}_Z$:
\begin{equation}
{\cal M}_Z\to {\cal M}_Z^{(-)}= -\displaystyle\frac{G_F}{2\sqrt{2}}
\left (g_v^{(e)}\ell_{\mu} {\cal J}_{\mu,5}^{(nc)}+ 
g_a^{(e)}\ell_{\mu,5}{\cal 
J}_{\mu}^{(nc)} \right ).
\label{eq:eq5}
\end{equation}
The P-odd asymmetry in the scattering of 
longitudinally 
polarized electrons can be written as:
\begin{equation} 
{A}=\displaystyle\frac{N_+-N_-}{N_++N_-}=
-\displaystyle\frac{G_F|k^2|}{2\sqrt{2}\pi\alpha}\displaystyle\frac{W^-}{W^{(
em)
}},
\label{eq:eq6}
\end{equation}
with two different contributions to ${W^-}$:
\begin{equation} 
{W^-}= g_a^{(e)}\widetilde{W_1}+g_v^{(e)}\widetilde{W_2},
\label{eq:eq7}
\end{equation}
where $W^{(em)}$ is proportional to $\overline{|{\cal M}_\gamma|^2}$:
\begin{equation}
W^{(em)}=\ell_{\mu\nu}W^{(em)}_{\mu\nu},
\label{eq:eq8}
\end{equation}
\begin{equation}
W^{(em)}_{\mu\nu}=\overline{{\cal J}_{\mu}^{(em)}{\cal J}_{\nu}^{(em)*}},
\label{eq:eq9}
\end{equation}
\begin{equation}
\ell_{\mu\nu}=2\left(  k_{1\mu}k_{2\nu}+k_{1\nu}k_{2\mu}-g_{\mu\nu}k_{1}\cdot 
k_2\right ),
\label{eq:eq10}
\end{equation}
and the overline in Eq. (\ref{eq:eq9}) stands for the sum over the final nucleon 
polarizations and the average over the polarizations of the initial nucleon in 
the process $\gamma^*+N\to N+\pi$. The quantities $\widetilde{W_1}$ and 
$\widetilde{W_2}$ in Eq. (\ref{eq:eq7}) 
characterize the interference of the electromagnetic hadronic current ${\cal 
J}^{(em)}$ with the vector and axial parts of the weak neutral current:
\begin{equation}
\widetilde{W_1}=\ell_{\mu\nu}W_{\mu\nu}^{(v)},
\label{eq:eq11}
\end{equation}
\begin{equation}
W_{\mu\nu}^{(v)}= \displaystyle\frac{1}{2}
\overline{{\cal J}_{\mu}^{(em)}{\cal J}_{\nu}^{(nc)*}},
\label{eq:eq12}
\end{equation}
\begin{equation}
\widetilde{W_2}=\ell_{\mu\nu}^{(a)}W_{\mu\nu}^{(a)},
\label{eq:eq13}
\end{equation}
\begin{equation}
W_{\mu\nu}^{(a)}=\displaystyle\frac{1}{2}
\overline{{\cal J}_{\mu}^{(em)}
{\cal J}_{\nu,5}^{(nc)*}},
\label{eq:eq14}
\end{equation}
and
\begin{equation}
\ell_{\mu\nu}^{(a)}=2i\epsilon_{\mu\nu\alpha\beta}k_{1\alpha}k_{2\beta},
\label{eq:eq15}
\end{equation}
where $\epsilon_{\mu\nu\alpha\beta}$ is the usual antisymmetric tensor.

In the product of the tensors $\ell_{\mu\nu}^{(a)}$ and $W_{\mu\nu}^{(a)}$, only 
the antisymmetric part of 
$W_{\mu\nu}^{(a)}$ contributes, whereas the quantity $\widetilde{W_1}$, 
Eq. (\ref{eq:eq11}),  
is determined by the symmetrical part of the tensor $W_{\mu\nu}^{(v)}$.

According to  Eq. (\ref{eq:eq4}), we can 
neglect the $\widetilde{W_2}$ contribution (the second P-odd contribution, which 
is 
induced by the axial part of the neutral weak current, is more model dependent 
and it will be the object of a detailed analysis in a 
subsequent paper).

In this approximation, the P-odd asymmetry is solely determined by the vector 
part of the hadronic neutral weak current:
\begin{equation}
{A}=-\displaystyle\frac{G_F|k^2|}{2\sqrt{2}\pi\alpha} 
\displaystyle\frac{\widetilde{W_1}}{W^{(em)}},
\label{eq:eq16}
\end{equation}
In order to calculate the ratio ${\widetilde{W_1}}/W^{(em)}$, we shall use 
the 
isotopic 
structure of the vector neutral current, which holds in the SM  when 
neglecting the 
contributions of the isoscalar quarks ($s$, $c$,...):
\begin{equation}
{\cal J}_{\mu}^{(nc)}=2 {\cal J}_{\mu}^{(1)}-4\sin^2\theta_W{\cal 
J}_{\mu}^{(em)}=
2(1-2sin^2\theta_W){\cal J}_{\mu}^{(em)}-2 {\cal J}_{\mu}^{(0)},
\label{eq:eq17}
\end{equation}
where 
\begin{equation}
{\cal J}_{\mu}^{(em)}={\cal J}_{\mu}^{(0)}+{\cal J}_{\mu}^{(1)},
\label{eq:eq18}
\end{equation} and 
${\cal J}_{\mu}^{(0)}$ and ${\cal J}_{\mu}^{(1)}$ are the isoscalar and 
isovector 
components of the electromagnetic hadronic current.
Considering the specific isotopic structure of ${\cal J}_{\mu}^{(nc)}$, Eq. 
(\ref{eq:eq17}),
the asymmetry ${A}$ for any process $\vec e+N\to e+N+\pi$ can be written 
as:
\begin{equation}
{A}=-\displaystyle\frac{G_F|k^2|}{2\sqrt{2}\pi\alpha}
\left [ 1-2\sin^2\theta_W +\Delta^{(s)}\right ],
\label{eq:eq19}
\end{equation}
where the quantity $\Delta^{(s)}$ results from the interference of the 
isoscalar 
component ${\cal J}_{\mu}^{(0)}$ of the electromagnetic current  with the 
full 
electromagnetic current in ${\cal J}_{\mu}^{(em)}$ i.e.:
\begin{equation}
\Delta^{(s)}=\displaystyle\frac{W^{(0)}}{W^{(em)}},~~W^{(0)}=
-\ell_{\mu\nu}\overline{{\cal J}_{\mu}^{(em)}{\cal J}_{\nu}^{(0)*}}.
\label{eq:eq20}
\end{equation}
One can see from Eq. (\ref{eq:eq19}) that the isovector part of the 
electromagnetic current induces a definite 
contribution to the P-odd asymmetry ${A}$, which is model 
independent and can be predicted in terms of the fundamental constants 
$G_F$, $\alpha$ and 
$\sin^2\theta_W$. Note that this contribution depends only on 
the variable $k^2$. Therefore, for reactions such as  
$ e^-+N\to e^-+\Delta$, $ e^-+d\to e^-+d+\pi^0$, where the electromagnetic 
current is pure isovector (and therefore $\Delta^{(s)}=0$), the asymmetry can be 
predicted exactly:
\begin{equation}
{A}=-\displaystyle\frac{G_F|k^2|}{2\sqrt{2}\pi\alpha}\left [ 
1-2\sin^2\theta_W 
\right ],
\label{eq:eq21}
\end{equation}
in agreement with ref.\cite{CG78} and neglecting the small contributions from 
the axial hadronic current, which is 
not 
considered 
here (note that $\displaystyle\frac{G_F}{2\sqrt{2}\pi\alpha}= 1.8\cdot 
10^{-4}$).
In particular, for the reaction  
$e^-+p\to e^-+\Delta^+$ this 
model-independent estimate of ${A}$  together with the possibility of a 
precise measurement of the P-odd asymmetry, open new ways to look for  
new physics \cite{Mu94} and to study effects due to the axial 
current.

In the next section, we will show that the quantity $\Delta^{(s)}$, in the 
near-threshold 
region for $e^-+N\to e^-+N+\pi$, as well as in the region of the 
$\Delta$ excitation, can be considered as a small correction to the main 
isovector contribution. Therefore, the uncertainty in the estimate of
$\Delta^{(s)}$ will affect very little the results.

Note that there is a model independent relation 
between the isoscalar components of the electromagnetic currents,
for the considered processes $\gamma^*+p\to n+\pi^+$ and
$\gamma^*+p\to p+\pi^0$, which holds for any interaction mechanism:
$${\cal J}_{\mu}^{(s)}(\gamma^*p\to n\pi^+)=
-\sqrt{2} {\cal J}_{\mu}^{(s)}(\gamma^*p\to p\pi^0).$$
>From Eq. (\ref{eq:eq19}) it appears that the  inclusive asymmetry ${A}$ 
depends on 
the 
variables $E_1$ and $W$  only through  the correction $\Delta^{(s)}$:
$$\Delta^{(s)}=\Delta^{(s)}(k^2,W,E_1).$$
Taking into account the longitudinal and transversal polarizations of the 
virtual 
$\gamma$ 
and $Z$-boson, the following representation for the correction $\Delta^{(s)}$ 
can be 
written (in case of a single channel: $e+p\to e+p+\pi^0$ or $e+p\to 
e+n+\pi^+$):
\begin{equation}
\Delta^{(s)}=\displaystyle\frac
{\sigma_T^{(s)}+\epsilon\displaystyle\frac{(-k^2)}
{\widetilde{k_0^2}}~\sigma_L^{(s)} }
{\sigma_T+\epsilon\displaystyle\frac{(-k^2)}{\widetilde{k_0^2}}~\sigma_L},
\label{eq:eq22}
\end{equation}
$$\epsilon^{-1}=1-2\displaystyle\frac{(-\vec k^2)}{k^2} 
\tan^2\displaystyle\frac{\theta_e}{2},~~
{\widetilde{k_0}}=\displaystyle\frac{W^2+k^2-m^2}{2W}, 
$$
where $\sigma_T(k^2,W)$ and $\sigma_L(k^2,W)$ are the total cross sections of 
virtual photon absorption in $\gamma^*+N\to N+\pi$:
$$\sigma_L=\int \overline{
\left |
{ \cal J}_z^{(em)} \right |^2} d\Omega_{\pi},$$
\begin{equation}
\sigma_T=\int \left ( 
\overline {\left |{\cal J}_x^{(em)}\right |^2 + \left |{\cal 
J}_y^{(em)}\right 
|^2 } 
\right )d\Omega_{\pi},
\label{eq:eq23}
\end{equation}
$d\Omega _{\pi}$ being the element of solid angle of the produced pion (in 
the 
CMS 
of the 
process $\gamma^*+N\to N+\pi$). We use here a coordinate system in which the 
z-axis is
along the 
three momentum of the virtual photon, and ${ \cal J}_x^{(em)}$, 
${ \cal J}_y^{(em)}$ and ${ 
\cal J}_z^{(em)}$ are the space components of the hadronic electromagnetic 
current.

The interference contributions $\sigma_L^{(s)}$ and  $\sigma_T^{(s)}$ are 
defined as 
follows:
$$
\sigma_L^{(s)}(k^2,W)=\int d\Omega_{\pi} {\cal R}e~ 
\overline{{\cal J}_z^{(em)}{\cal J}_z^{(0)*}},
$$
\begin{equation}
\sigma_T^{(s)}(k^2,W)= \int d\Omega_{\pi} {\cal R}e ~ \left [
\overline { { \cal J}_x^{(em)}{ \cal J}_x^{(0)*} +  {\cal J}_y^{(em)}
{\cal J}_y^{(0)*}}\right ],
\label{eq:eq24}
\end{equation}
where $\vec { \cal J}^{(0)} \left( { \cal J}_x^{(0)},  { \cal J}_y^{(0)}, 
{ \cal J}_z^{(0)}\right )$ are the space components of the isoscalar part of 
the 
hadronic 
electromagnetic current.

The lines above the products of the components of the electromagnetic 
currents 
mean 
the sum 
over the polarizations of the final nucleons and the average over the 
polarizations 
of the 
initial nucleons.

The inclusive asymmetry for $p(\vec e,e')N\pi$ with the contribution of two 
channels 
$p+\pi^0$ and $n+\pi^+$ in the final state, is determined by the following 
expressions:
$$ 
{A}=-\displaystyle\frac{G_F|k^2|}{2\sqrt{2}\pi\alpha}\left [ 
1-2\sin^2\theta_W 
+\Delta^{(s)}_{incl}\right ],$$
\begin{equation}
\Delta ^{(s)}_{incl}= \displaystyle\frac{
\Delta^{(s)}(\gamma^*p\to n\pi^+)+ 
R\Delta^{(s)}(\gamma^*p\to p\pi^0) }{(1+R)},
\label{eq:eq25}
\end{equation}
with
$$R=\displaystyle\frac{
\sigma_T(\gamma^*p\to p\pi^0)+\epsilon 
     \displaystyle\frac{
     (-k^2)}{\widetilde{k_0^2}}
     \sigma_L(\gamma^*p\to p\pi^0)
     }
{\sigma_T(\gamma^*p\to n\pi^+)+\epsilon 
\displaystyle\frac{(-k^2)}{\widetilde{k_0^2}}
\sigma_L(\gamma^*p\to n\pi^+)}.$$
Therefore, the P-odd inclusive asymmetry ${A}$ for 
$p(\vec e,e')N\pi,~N\pi=(p+\pi^0)+(n+\pi^+)$ is determined by a set of four 
total cross sections:
$$\sigma_T(k^2,W),~\sigma_L(k^2,W), ~\sigma_T^{(s)}(k^2,W), 
~\mbox{and}~\sigma_L^{(s)}(k^2,W),$$
for each $\gamma^*+p\to n+\pi^+$ and 
$\gamma+p\to p+\pi^0$ processes (8 in total), as functions of two independent 
kinematical variables 
$k^2$ 
and $W$. 
The polarization parameter $\epsilon$, $0\le \epsilon \le 1$, which 
represents the linear 
polarization of the virtual photon, contains the dependence on the 
kinematical 
conditions of the electrons in the initial and final states (i.e. initial energy 
and scattering angle).

In the present calculation we shall use the following parametrization of the 
spin 
structure of the matrix element for $\gamma^*+N\to N+\pi$, in terms of six 
standard contributions:
$${\cal M}(\gamma^*N\to N\pi)=\chi_2^\dagger {\cal F}\chi_1,$$
\begin{equation}
{\cal F}=i\vec e\cdot\hat{\vec k}\times\hat{\vec q} f_1+
\vec\sigma\cdot\vec e f_2 +\vec\sigma\cdot\hat{\vec k}~
\vec e\cdot\hat{\vec q}f_3+\vec\sigma\cdot\hat{\vec q}~\vec e\cdot\hat{\vec 
q}f_4
\label{eq:eq26}
\end{equation}
$$+\vec e\cdot\hat{\vec k}(\vec\sigma\cdot\hat{\vec 
k}f_5+\vec\sigma\cdot\hat{\vec q}f_6),$$
where $\chi_1$ and $\chi_2$ are the two-component spinors of the initial and 
final nucleons, $\vec e$ is the three-vector of the virtual photon 
polarization, $\hat{\vec k}$ and $\hat{\vec q}$ are the unit vectors along the 
3-momentum of the $\gamma^*$ and $\pi$ in the  CMS of the $\gamma^*+N\to 
N+\pi$- 
reaction.
The complex scalar amplitudes $f_i$, which are functions of three independent 
kinematical variables, $f_i=f_i(k^2,W,cos\theta_\pi)$, can be related to the 
usual set of amplitudes, $F_i$, when the operator 
$\vec\sigma\cdot\hat{\vec q}~\vec\sigma\cdot\vec e\times\hat{\vec k}$ is used 
instead of $i\vec e\cdot\hat{\vec k}\times\hat{\vec q}$:
$$f_1=F_1,$$
$$f_2=F_1-cos\theta_{\pi}F_2,$$
$$f_3=F_2+F_3,$$
$$f_4=F_4.$$

The results of averaging over the polarization states of the initial nucleon 
and 
summing over the polarizations of the produced nucleon gives:
$$\overline{|{\cal J}_x^{(em)}|^2+|{\cal J}_y^{(em)}|^2}=2|f_2|^2+
\sin^2\theta_{\pi}(|f_1|^2+|f_3|^2+|f_4|^2+$$
$$+2{\cal R}e (f_2f_4^*+ \cos \theta_{\pi} f_3f_4^*)),$$
$$\overline{|{\cal J}_z^{(em)}|^2}=|f_2+\cos \theta_{\pi} f_3+f_5|^2
+|\cos \theta_{\pi}f_4+f_6|^2
$$
\begin{equation}
+2\cos \theta_{\pi}{\cal R}e(f_2+\cos \theta_{\pi}f_3+f_5)
(\cos \theta_{\pi}f_4+f_6)^*.
\label{eq:eq27}
\end{equation}
To calculate the cross sections $\sigma_T^{(s)}$ and $\sigma_L^{(s)}$, in 
Eqs. (\ref{eq:eq27}) the following substitutions are made:
$$
|f_i|^2\to {\cal R}e f_if_i^{(s)},
$$
\begin{equation}
2{\cal R}e f_if_j^*\to 
{\cal R}e (f_if_j{^{(s)}}^*+f_jf_i{^{(s)}}^*),~i,j=1,..6,
\label{eq:eq28}
\end{equation}
where $f_i^{(s)}$ are the scalar amplitudes, describing the isoscalar part of 
the hadronic electromagnetic current ${\cal J}_\mu^{(em)}$.

\section{Model for $\lowercase{e^-}+N \to \lowercase{e^-}+N+\pi$}

We use here the standard approach for the calculation of the electromagnetic 
current for the $\gamma^* +N\to N+\pi$ processes, which  describes 
satisfactorily well the existing photo- and 
electro-production data, in the region of $W$ starting from 
threshold, $W=m+m_\pi$, up to $W\simeq$ 1.3 GeV (the $\Delta$ 
excitation region). This approach takes into account the following three 
contributions:
\begin{itemize}
\item Born terms in the $s$, $t$ and $u$ channels, 
\item vector meson ( $\omega$ and $\rho$) exchanges in the $t$-channel,
\item $\Delta$-isobar excitation in the $s$ channel.
\end{itemize}
Using the isotopic structure of the 'strong' vertices on the diagrams (Fig. 
2), the scalar 
amplitudes for each $\gamma^*+N\to N+\pi$ process can be written as:
\begin{equation}
f_i=\sqrt{(E_1+m)(E_2+m)}\left [ a_s f_{i,s}+a_uf_{i,u}+a_t f_{i,t} 
+a_{\rho}f_{i,\rho}+a_{\omega}f_{i,\omega}+a_{\Delta}f_{i,\Delta}\right ],
\label{eq:eq29}
\end{equation}
where $f_{i,s}...f_{i,\Delta}$ characterize the contributions of the 
different 
Feynmann diagrams to the scalar amplitudes $f_i$, $i=1-6$. The energies $E_1$ 
and $E_2$ of the initial and final nucleons are determined by the following 
formulae:
$$E_1=\displaystyle\frac{s+m^2-k^2}{2\sqrt{s}},~~
E_2=\displaystyle\frac{s+m^2-m_{\pi}^2}{2\sqrt{s}}.$$
The isotopic numerical coefficients $a_s...a_{\Delta}$ for the two  
processes $\gamma^*+p\to p+\pi^0$ and 
$\gamma^*+p\to n+\pi^+$ are shown in Table 1.

The scalar amplitudes for the isoscalar part of the 
electromagnetic current can be written as follows:
$$f_i^{(s)}(\gamma^*p\to p \pi^0)=-
\left (f_{i,s}^{(s)}+f_{i,u}^{(s)}+f_{i,\rho}\right ) 
\sqrt{ (E_1+m)(E_2+m)},$$
where $f_{i,s}^{(s)}$ and $f_{i,u}^{(s)}$ are the amplitudes for the 
isoscalar 
part of the $s$- and $u$- channel Born diagrams. 

One can see now that, in the framework of the considered approach, the main 
contributions to ${\cal 
J}_\mu^{(em)}$ have an isovector nature:
\begin{itemize}
\item $\Delta$-excitation in $\pi^+$ and $\pi^0$ production,
\item $\pi^+$-exchange for $\pi^+$ production,
\item $\omega$-exchange for $\pi^0$ production,
\item contact term for $\pi^+$ production (in the case of a pseudovector 
$\pi NN$-interaction),
\item $s+u$ Born contributions.
\end{itemize}
Therefore, the isoscalar electromagnetic current can only contain 
 the following contributions:
\begin{itemize}
\item $\rho-$exchange for $\pi^0$ and $\pi^+$-production,
\item the isoscalar part of the $s+u$-diagrams.
\end{itemize}

However these isoscalar contributions are  small in comparison with the 
corresponding isovector ones. Indeed, the $\rho$-exchange 
term is smaller than the $\omega$ -exchange term, due to the following 
reasons:
\begin{itemize}
\item $g_{\rho\pi\gamma}\simeq \displaystyle\frac{1}{3}g_{\omega\pi\gamma}$: 
suppression at electromagnetic vertices; 
\item $g_{\rho NN}\simeq \displaystyle\frac{1}{6}g_{\omega NN}$: suppression 
at 
the 
strong vertex.
\end{itemize}
In the same way, the isoscalar Born contribution due to the nucleon magnetic 
moment, for 
example, is smaller than the isovector contribution:

$$\displaystyle\frac{|\mu_p+\mu_n|}{|\mu_p-\mu_n|}=\displaystyle\frac{|1.79-1
.91
|}{1.79+1.91}\approx 10^{-2}$$
This clearly shows that $\Delta^{(s)}$ can be considered a small correction 
to 
the 
model-independant prediction of Eq. (\ref{eq:eq21}).
Let us briefly discuss now the properties of the suggested model, for the 
 $\gamma^*+p\to N+\pi$ processes.

\subsection{Born contribution: $s-$channel}

Using a pseudoscalar $\pi NN$-interaction, we can write the relativistic 
invariant expression for the matrix element of the $\gamma+p \to n+\pi^+$ 
reaction in the following form:
$${\cal M}_B=eg({\cal M}_s+{\cal M}_u+{\cal M}_t) ,$$
$${\cal M}_s=\overline{u}(p_2)\gamma_5\displaystyle\frac{\hat{p_2}+
\hat{q}+m}{s-m^2}\left(F_{1p}\hat 
e+F_{2p}\displaystyle\frac{\sigma_{\mu\nu}e_{\mu}k_{\nu}}{2m}\right 
)u(p_1),$$
$${\cal M}_u=\overline{u}(p_2)\left(F_{1n}\hat 
e+F_{2n}\displaystyle\frac{\sigma_{\mu\nu}e_{\mu}k_{\nu}}{2m}\right )
\displaystyle\frac{\hat{p_2}+
\hat{q}-m}{u-m^2}\gamma_5 u(p_1),$$
$${\cal M}_t= \displaystyle\frac
{(2e\cdot q-e\cdot k)}{t-m^2_{\pi}}\overline{u}(p_2)\gamma_5 u(p_1),$$
where $s$, $t$, and $u$ are the standard Mandelstam variables:
$$s=(p_2+q)^2,~~t=(p_1-p_2)^2,~~u=(p_2-k)^2,$$
$k$, $q$, $p_1$ and $p_2$ are the four-momenta of $\gamma^*$, $\pi$, initial 
and final nucleons, $e$ is the four-vector of the virtual photon 
polarization, 
$g$ is the $\pi NN$ coupling constant (for a pseudoscalar interaction), 
$F_{1p}(k^2)$ 
and $F_{2p}(k^2)$ ($F_{1n}(k^2)$ and $F_{2n}(k^2)$) are the Dirac and Pauli 
electromagnetic form factors of the proton (neutron). The 
electromagnetic form factor of the nucleon is usually parametrized in form of 
a
$k^2$-dependence of the electric ($G_{EN}$) and magnetic ($G_{MN}$) nucleonic 
form factors:
$$F_{1N}(k^2)=\displaystyle\frac{G_{EN}(k^2)-\tau G_{MN}(k^2)}{1-\tau },$$
$$F_{2N}(k^2)=\displaystyle\frac{-G_{EN}(k^2)+G_{MN}(k^2)}{1-\tau },~~\tau 
=\displaystyle\frac{k^2}{4m^2}.$$

A simple dipole dependence of $G_{Ep}$, $G_{Mp}$ and $G_{Mn}$:
$$G_{Ep}(k^2)=G_{Mp}(k^2)/\mu_p=G_{Mn}(k^2)/\mu_n=
 \displaystyle\frac{1}
 {\left [1-\displaystyle\frac{k^2}{0.71 (GeV/c)^2}\right ]^2},$$
with $\mu_p=2.79, \mu_n=-1.91,$ has been considered a good parametrization of 
the existing experimental data, 
in 
a wide region of space-like momentum transfer while $G_{En}(k^2)=0$.
However a very recent direct measurement \cite{Cfp99} of the ratio 
$G_{Ep}/G_{Mp}$ 
shows some deviation of $G_{Ep}$ from a dipole behavior, in 
the region $0\le -k^2\le 3.5$ (GeV/c)$^2$. This high precision experiment is 
based 
on the measurement of the polarization of the final protons in $\vec e+p \to 
e+\vec p$, in the elastic scattering of longitudinally polarized electrons 
\cite{Re68}.

This effect should be taken into account in future calculations, as well as 
the fact that $G_{En}$ deviates from zero, at least in the region $k^2\le 
1(GeV/c)^2$. The last direct measurement of $G_{En}$, 
in $\vec e +\vec d\to e+X$ \cite{GEnNikhef}
confirms some previous estimates of the neutron form 
factor based on the Saclay  $ed$ elastic scattering  data \cite{Pl90}.

In the Vector Dominance Model (VDM) approach, the pion electromagnetic form 
factor $F_{\pi}(k^2)$ is 
described by:
$$F_{\pi}(k^2)=\left ( 1-\displaystyle\frac{k^2}{{m_\rho}^2}\right )^{-1},$$
where $m_{\rho}$ is the $\rho$-meson mass.

Note that the electromagnetic current for the reaction $\gamma^*+p\to 
p+\pi^0$, 
corresponding to the sum of the Born diagrams in the $s$ and $u$-channels, is 
conserved for any form 
factors $F_{1p}$ and $F_{2p}$ in the whole kinematical region. This is not 
the 
case for the reaction $\gamma^*+p\to n+\pi^+$ (\ref{eq:eq26}), as one can show 
that the divergence of the corresponding 
electromagnetic current (in the Born approximation, for the sum of the $s$, $t$, 
and 
$u$-contributions) is proportional to the following combination of the 
electromagnetic form factors:
$$k\cdot {\cal J}^{(B)}(\gamma^* p\to n\pi^+)= e g \sqrt{2}\left 
(F_{1p}- F_{1n}- F_{\pi}\right ) \overline{u}(p_2)\gamma_5 u(p_1).$$
The simplest way to conserve the hadronic electromagnetic current, is to extend 
to all values of $k^2$ the following relation:
\begin{equation}
F_{1p}(k^2)=F_{1n}(k^2)+F_{\pi}(k^2),
\label{eq:eq30}
\end{equation}
which is in general valid only for $k^2=0$. However existing data on pion and 
nucleon form factors are in contradiction with relation 
(\ref{eq:eq30}). A possible way to avoid this difficulty is to renormalize 
the 
matrix element ${\cal M}_B(\gamma^*p\to n\pi^+)$ in the following way:
\begin{equation}
{\cal M}_B\to {\cal M}'_B={\cal M}_B+eg\displaystyle\frac {e\cdot k}{k^2}
\overline{u}(p_2)\gamma_5 u(p_1) \left (-F_{1p}+F_{1n}+F_{\pi}\right ),
\label{eq:eq31}
\end{equation}

The electromagnetic current, corresponding to the new Born matrix element 
${\cal M}'_B$ is conserved for any form factor. Note that such a procedure 
changes only
$\sigma_L$, without any effect on the transversal cross-sections 
$\sigma_T(k^2,W)$ and $\sigma_T^{(s)}(k^2,W)$. In our considerations we shall 
use the procedure (\ref{eq:eq31}), while  relation 
(\ref{eq:eq30}) was taken in ref.\cite{Va95}.

The scalar amplitudes $f_i$, corresponding to different diagrams of the Born 
mechanism, are given in the Appendix.
\subsection{Vector meson exchange}
The matrix element ${\cal M}_V$, corresponding to vector meson exchange 
in the $t-$channel can be written in the 
following form:
\begin{equation}
{\cal M}_V=\displaystyle\frac {eg_{V\pi\gamma^*}(k^2)}{t-m_V^2}
\epsilon_{\mu\nu\alpha\beta}e_{\mu}k_{\nu}{\cal J}_{\alpha}^{(V)}q_{\beta},
\label{eq:eq32}
\end{equation}
$$ {\cal J}_{\alpha}^{(V)}=\overline{u}(p_2)\left [ \gamma_{\alpha}F_1^V(t) - 
\displaystyle\frac{F_2^V(t)}{2m}\sigma_{\alpha\beta}(p_1-p_2)_{\beta}\right ] 
u(p_1) $$
where $g_{V\pi\gamma^*}(k^2)$ is the electromagnetic form factor for the 
$V\pi\gamma^*$-vertex, $m_V$ is the vector meson mass,
$F_1^V(t)$ and $F_2^V(t)$ are the "strong" form factors for the $V^*NN$ 
vertex 
 (with a virtual V-meson). In principle the "static" values of these form 
factors (i.e. for $t=0$), are related to the $\omega NN$ and $\rho NN$ coupling 
constants:
$$F_1^V(0)=g_{VNN},~~F_2^V(0)/F_1^V(0)=\kappa_V.$$
An estimate for the $\omega NN$ coupling constants, based on the 
Bonn potential \cite{Ma87}, gives:
$$\displaystyle\frac{g_{\omega NN}^2}{4\pi}=20,~~\kappa_{\omega}=0$$
The $\rho NN$ coupling constants can be estimated 
from pion photoproduction data \cite{Va95}:
$$\displaystyle\frac{g_{\rho NN}^2}{4\pi}=0.55,~~\kappa_{\rho}=3.7$$
Note that the constants $\kappa_V$ can be identified in VDM, with the values of 
the 
isoscalar and isovector anomalous magnetic moment of the nucleon.
The VDM allows to write the following parametrization for 
the 
$k^2$ dependence of the electromagnetic form factor for the $\gamma^*+V\to \pi$ 
vertex: 
$$g_{V \pi\gamma^*}(k^2)=
\displaystyle\frac{g_{V\pi\gamma}(0)}{1-k^2/m_V^2},$$
where $m_V$ is the mass of the $\rho$ or $\omega$ vector meson.

The $g_{V\pi\gamma}(0)$ coupling constant can be fixed by the width of the 
radiative 
decay $V\to \pi\gamma$, through the following formula:
$$\Gamma(V\to\pi\gamma)=\displaystyle\frac{\alpha}{24}|g_{V\pi\gamma}(0)|^2
\left (1-\displaystyle\frac{m_{\pi}^2}{m_V^2}\right )^3.$$
The numerical 
estimate, is based on  the following values 
\cite{pdg}:
$$Br(\omega\to 
\pi^0\gamma)=\Gamma(\omega\to\pi\gamma)/\Gamma_{\omega}=(8.5\pm 
0.5) 10^{-2},$$
$$\Gamma_{\omega}=(8.41\pm 0.09) ~\mbox{MeV},$$
$$Br(\rho^0\to \pi^0\gamma)=(6.8\pm 1.7)\cdot 10^{-4},$$
$$Br(\rho^{\pm}\to \pi^{\pm}\gamma)=(4.5\pm 0.5)\cdot 10^{-4},$$
$$\Gamma_{\rho}=(150.7\pm 1.1) ~\mbox{MeV}.$$
The following relation 
holds for the hadronic electromagnetic current, :
$$Br(\rho^{\pm}\to \pi^{\pm}\gamma)=Br(\rho^0\to \pi^0\gamma),$$
Therefore any violation of this relation is an indication of the presence 
of an isotensor component of the electromagnetic current, which is absent, 
however, 
at 
the quark level.
So, a precise experiment with the simultaneous determination of the two 
coupling constants for $\rho^0\to\pi^0\gamma$ and 
$\rho^{\pm}\to\pi^{\pm}\gamma$ would be very important. It would not 
only constitute a test of the isotopic properties of the hadronic 
electromagnetic current, but also have application in the 
calculation of the meson exchange current contributions (MEC) to the deuteron 
electromagnetic form factors. Such contributions are considered to be very 
important at high momentum transfer.

Note in this respect that the $t$-dependence of the form factors 
$F_1^{\omega}(t)$ and $F_2^{\omega}(t)$ is also important, for the correct 
calculation of MEC, in case of elastic electron-deuteron scattering.  
Moreover, the relative sign of the $V-$exchange and Born contributions to the
$\gamma^*+p\to N+\pi$ processes, is generally not known. So, we shall 
consider 
here both 
relative signs. The determination of this sign is also important for MEC 
calculations, because in both cases ($e+d\to e+d$ and $\gamma^*+p\to N+\pi$) 
the 
set of constants and form factors are the same.

Finally we stress that the electromagnetic current, corresponding to  
vector 
meson exchange in the processes $\gamma^*+p \to N+\pi$ is automatically 
conserved, independently of the parametrization of the strong form factors 
$F_1(t)$ and  $F_2(t)$ and the electromagnetic form factor 
$g_{V\pi\gamma^*}(k^2)$.

\subsection{$\Delta$-excitation}
This contribution can be analyzed in a relativistic framework \cite{HD95}, 
considering a 
virtual $\Delta$ as a Rarita-Schwinger field with spin 3/2 but in this approach 
it is difficult to treat off-shell effects. First of all, this means that 
$\Delta-$exchange may contain contributions from a state with spin 1/2 as 
well 
as antibaryonic terms with negative P-parity and s=1/2 and 3/2. Therefore the 
description of the $\Delta-$isobar, with ${\cal J}^P=3/2^+$, especially in 
the 
$s-$channel 
is not straightforward. To avoid these complications, we choose here a direct 
parametrization of the $\Delta $ contribution. Note that the CMS for 
$\gamma^*+p\to\Delta^+\to N+\pi$ is the optimal frame, because the 
three-momentum of the $\Delta$ is zero, so that the $ \Delta$ can be 
described 
by a 
two-component spinor, with a vector index, $\vec\chi$, which satisfies the 
following auxiliary condition:
$$\vec\sigma\cdot\vec\chi=0,$$
typical for a pure spin 3/2 state. Using this condition, it is possible to 
find 
the following expression for the $\Delta$-density matrix:
\begin{equation}
\rho_{ab}=\overline{\chi_a\chi_b^\dagger}=\displaystyle\frac{2}{3}
(\delta_{ab}-
\displaystyle\frac{i}{2}\epsilon_{abc}\sigma_c),
\label{eq:eq33}
\end{equation}
with the normalization condition: $Tr \rho_{\alpha\alpha}=2s_{\Delta}+1=4$.

In this formalism the $\Delta N\pi$-vertex can be parametrized as follows:
\begin{equation}
{\cal M}_{\Delta N\pi}=g_{\Delta N\pi} \chi^\dagger\vec\chi\cdot\hat{\vec 
q},
\label{eq:eq34}
\end{equation}
where $\chi$ is the 2-component spinor of the nucleon in the decay $\Delta\to 
N+\pi$, $\hat{\vec q}\:$  is the unit vector along the pion three momentum, in 
the $\Delta$ rest frame, and the constant $g_{\Delta N\pi}$ characterizes the 
width of the strong decay $\Delta\to N+\pi$.

Taking into account the conservation of the total angular momentum and of 
the P-parity in the electromagnetic decay $\Delta\to N+\gamma$ with production 
of M1 photons, the following 
expression can be written for the matrix element:
\begin{equation}
{\cal M}_{\Delta N\gamma}=eg_{\Delta N\gamma} 
\chi^\dagger\vec\chi\cdot\vec e\times\hat{\vec k},
\label{eq:eq35}
\end{equation}
where $g_{\Delta N\gamma}$ is the constant of the magnetic dipole radiation 
(or the magnetic moment for the transition $\Delta\to N+\gamma$), $\vec e$ and 
$\hat{\vec k}$ are the photon polarization three-vector and unit momentum 
vector 
along 
the three-momentum of $\gamma$, respectively.

In the general case, the transition $\gamma^*+N\to \Delta$ must be described 
by 
three different form factors, corresponding to the absorption of $M1$, 
$E2_{t}$ 
(transversal) and $E2_{\ell}$ (longitudinal) virtual photons. But the 
existing 
experimental data about pion photo and electro-production of pions on the 
nucleons (in the $\Delta$ resonance region) indicate that the $M1$ term is 
dominant \cite{Fr99}, therefore in our analysis we will consider only this 
form 
factor.


Using expressions (\ref{eq:eq34}) and (\ref{eq:eq35}) for both the $\Delta 
N\pi$ and $\Delta N\gamma$ vertices, one can write the matrix element of the 
$\Delta-$contribution in the $s-$channel (Fig. 2e) as follows:
\begin{equation}
{\cal M}_{\Delta N\gamma }=\displaystyle\frac{eG(k^2)|\vec q|}
{M_{\Delta}^2-s-i\Gamma_{\Delta} M_{\Delta}} \sqrt{(E_1+m)(E_2+m)} 
\chi_2^\dagger( 
2\delta_{ab}-i\epsilon_{abc}\sigma_c)\chi_1\hat {q_a}(\vec 
e\times\hat{\vec k})_b,
\label{eq:eq36}
\end{equation}
where $ M_{\Delta}$ and $\Gamma_{\Delta}$ are the mass and width of $\Delta$ and 
$G(k^2)$ is 
proportional to the magnetic form factor of the $\gamma^*+N\to \Delta$ 
transition.

The following $\Delta$ contributions to the scalar amplitudes, 
$f_{i}^\Delta$
can be derived:
$$f_{1}^\Delta=2\Pi(s,k^2)$$
$$ f_{2}^\Delta=\cos \theta_{\pi}\Pi(s,k^2),$$
$$ f_{3}^\Delta=-\Pi(s,k^2),$$
$$ f_{4}^\Delta=f_{5}^\Delta=f_{6}^\Delta=0,$$
where we use the notation:
$$\Pi(s,k^2)=\displaystyle\frac{ G(k^2)|\vec q|}{M_{\Delta}^2-s-i\Gamma_{\Delta} 
M_{\Delta}}.$$
The role of the factor $\vec q$ is to correctly  describe the threshold 
behavior 
of the $M1$ amplitude for $\gamma+p\to N+\pi$, according to the P-wave nature 
of 
the produced pion.
We shall use the following formula for the $k^2$-dependence of the transition 
electromagnetic form factors:
\begin{equation}
G(k^2)=\displaystyle\frac{G(0)} {\left (1-\displaystyle\frac{k^2}
{0.71~(GeV/c)^2}\right )^2\left (1-\displaystyle\frac{k^2}{ m_x^2}\right )}.
\label{eq:eq37}
\end{equation}
The factor $(1-\displaystyle\frac{k^2}{m_x^2})^{-1}$, with $m_x=6$ GeV$^2$, is 
included in order to take into account a steeper decreasing of 
$G(k^2)$ in comparison with the dipole behavior of the elastic 
electromagnetic 
form factors of the nucleons \cite{Fr99}.\\

The normalization constant $G(0)$ can be found according to the following 
procedure. 
Using Eq. (\ref{eq:eq36}), let us calculate first the differential cross 
section for $\pi^0$-photoproduction:
\begin{equation}\displaystyle\frac{d\sigma}{d\Omega}
(\gamma p\to p\pi^0)=
\displaystyle\frac{\alpha}{32\pi}
\displaystyle\frac{q_{\Delta}^3}{k_{\Delta}}
\displaystyle\frac{(E_{1\Delta}+m)(E_{2\Delta}+m)}{M_{\Delta}^4
\Gamma_{\Delta}^2} 
G^2(0)(5-3\cos^2\theta_\pi) \label{eq:eq38},
\end{equation}
at $s=M_{\Delta}^2$, where the $\Delta$-excitation in the $s-$channel is the 
main 
mechanism. So, our parametrization of the $\Delta$-contribution describes 
correctly 
the angular dependence $(5-3\cos^2\theta_\pi)$, typical for the magnetic 
excitation of a $\displaystyle\frac{3}{2}^+$ state in $\gamma+p\to \Delta^+\to 
N+\pi$. Therefore, the total cross section can be written as:
$$ \sigma_t(\gamma p\to p \pi^0)=
\displaystyle\frac{\alpha}{2}
\displaystyle\frac{q_{\Delta}^3}{k_{\Delta}}
\displaystyle\frac{(E_{1\Delta}+m)(E_{2\Delta}+m)}{M_{\Delta}^4
\Gamma_{\Delta}^2} 
G^2(0),$$
where:
$$E_{1\Delta}=\displaystyle\frac{M_{\Delta}^2+m^2}{2M_{\Delta}},
~~E_{2\Delta}=\displaystyle\frac{M_{\Delta}^2+m^2-m^2_{\pi}}{2M_{\Delta}},$$
$$k_{\Delta}=\displaystyle\frac{M_{\Delta}^2-m^2}{2M_{\Delta}},~~q_{\Delta}=
\sqrt{E_{2\Delta}^2-m^2}.$$
We can approximate with a good accuracy $\sigma_t(\gamma p\to p\pi^0)$ by a 
single $\Delta-$resonance contribution.
For a numerical estimate of 
$G(0)$ we use $\sigma_t\simeq 250\cdot 10^{-30}$ cm$^2$.

Note again that this procedure cannot determine the sign of $G(0)$. However 
for 
the 
$\gamma+p\to n+\pi^+$ reaction, there is a strong interference between the 
pion 
diagram 
and the $\Delta-$contribution. 
The comparison of the calculations using different signs
with the experimental $\theta_{\pi}$-dependence in the resonance region
allows to fix the corresponding relative sign. Two remarks should be done 
about this procedure:
\begin{itemize}
\item there is no ambiguity concerning off-mass shell effects for the 
$\Delta$-contribution, at least in the $s-$channel,
\item this special contribution is gauge invariant.
\end{itemize}
We neglect in our consideration the $\Delta$-exchange in the $u-$channel. The 
main 
reason to include this contribution is to have the crossing symmetry of the 
model. 
This is in principle an important property of the photoproduction amplitude, 
in 
particular in connection with the dispersion relation approach.
However in the framework of phenomenological approaches, this symmetry is 
strongly violated. For example, the $s-$channel $\Delta-$ contribution induces 
an amplitude which is mostly complex (with a typical Breit-Wigner behavior), 
whereas the $u$-channel contribution results in a real amplitude. The inclusion 
of different form factors for the $s-$ and $u-$channel violates the crossing 
symmetry, which is important for the Born contributions. 
This appears clearly for 
the reaction $\gamma+p\to p+\pi^0$, because here the crossing symmetry is 
correlated with the gauge invariance of the electromagnetic interaction, and 
the violation of the crossing symmetry has for direct consequence the violation 
of the current conservation. 
On the contrary, for the $\Delta$-contribution, this important correlation is 
absent.
\section{Numerical predictions and discussion}

Having determined all the parameters of the model, it is possible in principle 
to 
calculate all observables for the processes $e^-+N\to e^-+N+\pi$ (on proton 
and 
neutron targets) in the kinematical region from threshold to the
$\Delta$-resonance  region ($W\le 1300$ MeV), for any value of the pion 
production angle, $\theta_\pi$, (in the CMS of the $\pi N$-system,) and of the 
four momentum transfer $k^2$.

In order to test the present model, we used existing experimental data 
on the angular dependence of the differential cross sections for both 
 the $\gamma+p\to p+\pi^0$ and $\gamma+p\to n+\pi^+$ reactions. This 
comparison 
allowed to fix empirically the relative sign of the different contributions: 
Born, $\Delta$-excitation (in the $s-$channel) and vector meson exchange (in 
the 
$t-$ channel). The relative sign of all three diagrams for the Born 
approximation 
in the case of the process $\gamma+p\to n+\pi^+$ are fixed by gauge 
invariance, but it is necessary to find the relative signs between the Born 
amplitudes, on one side, and the $\Delta$-isobar  and vector meson 
exchange contributions, on another side. The  $\gamma+p\to n+\pi^+$ reaction 
is 
more sensitive 
to the signs of the $\Delta$-contribution and $\rho^+$-exchange. Then the 
data about the differential cross sections for $\gamma+p\to p+\pi^0$ allow a 
further check and give a constrain for the $\omega-$exchange amplitude.

Note that the sign of the $\rho-$exchange contribution relative to the Born 
contribution (in both the  $\gamma+p\to p+\pi^0$ and $\gamma+p\to 
n+\pi^+$  reactions) has to be the same as the relative sign 
of meson exchange currents (due to the $\rho\pi\gamma^*$ meson-exchange 
mechanism in the calculation of  the
electromagnetic form factors of the deuteron) with respect to the amplitude 
in the impulse approximation for elastic $ed-$scattering. This represents an 
important link between very different physical problems.

In order to obtain a good description of the experimental data for 
$\gamma+p\to 
p+\pi^0$ and $\gamma+p\to n+\pi^+$ we introduced small corrections to the 
different contributions. For the reaction $\gamma+p\to p+\pi^0$ a form factor 
was added to the Born contribution. The $u-$channel nucleon contribution for 
$\gamma+p\to n+\pi^+$ can be neglected without violating gauge invariance, 
because its magnetic content satisfies alone the current conservation 
condition. As a matter of fact this contribution has a diverging behavior at 
large angle, which is typically corrected by introducing an {\it ad-hoc} form 
factor. We choose to replace this contribution with a somewhat simplified
phenomenological (S-wave like) contribution: 
$a\left (1-\displaystyle\frac{t}{1.2}\right )
\displaystyle\frac{1.2\mbox{ GeV}}{W},$ 
where $a$ is a parameter which is adjusted in order to reproduce at best
the  $\pi^0$ photoproduction data.

We did not attempt to reproduce with a good accuracy the threshold 
behavior 
of the 
$\gamma+p\to p+\pi^0$ and $\gamma+p\to n+\pi^+$ amplitudes. A precise 
description of this behavior, in particular for the process 
$\gamma+p\to p+\pi^0$, can be obtained, for example, in the framework of the 
Chiral Perturbative Theory  approach \cite{VB96}. For inclusive calculations,  
a qualitative description of the data in the threshold region is sufficient.

The quality of our model is shown in Fig. 3, where we present the comparison of 
our predictions with the experimental data on the differential cross sections
for the $\gamma+p\to p+\pi^0$ and $\gamma+p\to n+\pi^+$ reactions, in the 
kinematical region 
where our model can be considered a reasonable approach. Indeed the 
unpolarized differential cross sections are well described. We did not apply the 
model to polarization observables. In particular different $T-$odd observables, 
such as, for example, the target asymmetry or the polarization of the final 
nucleons, are very sensitive to the relative phases of the different 
contributions. A good 
description requires a very precise treatment of the unitarity condition as 
well as of T-invariance of the hadron electromagnetic interaction, which are not 
so important for the differential or total cross section.

Therefore, after having determined the relative signs of the different 
contributions, our model can be generalized to pion electroproduction.
Our aim is the calculation of the inclusive P-odd asymmetry ${A}$, in 
$p(\vec e ,e ')X$, for the sum of two possible channels, $X=p+\pi^0$ and 
$X=n+\pi^+$. One can see, from Eq. (\ref{eq:eq25}), that such asymmetry is 
determined by the following ratios of inclusive cross sections:
$$R_L^{(s)}=\displaystyle\frac{
\sigma_L^{(s)}(k^2,W)
}{
\sigma_T(k^2,W)
},
~~R_T^{(s)}=\displaystyle\frac{
\sigma_T^{(s)}(k^2,W)
}{
\sigma_T(k^2,W)
},~~
~~R_{LT}=\displaystyle\frac{
\sigma_L(k^2,W)
}{
\sigma_T(k^2,W)},$$
for both channels, $\gamma^*+p\to p+\pi^0$ and $\gamma^*+p\to n+\pi^+$,
and $$R_{pn}=\displaystyle\frac{\sigma_T(\gamma^*p\to 
p\pi^0)}{\sigma_T(\gamma^*p\to 
n\pi^+)},$$
which characterizes the relative role of the two channels.
The 2-dimensional plots of these ratios as functions of $k^2$ and $W$ are 
shown in Fig. 4 and 5, for  the reactions 
$\gamma^*+p\to p+\pi^0$ and $\gamma^*+p\to n+\pi^+$, respectively. 

For $\pi^0$-electroproduction, both $R_L^{(s)}$ and $R_T^{(s)}$ are small 
corrections to ${A}$. In the considered kinematical region, they are 
positive and tend to 
decrease in the region of the $\Delta$ resonance, due to the dominance of the 
isovector resonance contribution. The behavior of all these ratios in the 
threshold region can be 
improved, as we discussed above.

In the case of the $\gamma^*+p\to n+\pi^+$ reaction, the corresponding 
corrections 
are also small, especially  $R_L^{(s)}$.
Note that  $R_T^{(s)}$ is negative 
in the whole region of $k^2$ and $W$.

Combining these results it is possible to calculate the resulting asymmetry 
${A}$ for the sum of both channels, again in a 2-dimensional 
representation (Fig. 6). The dependence on the detailed electron kinematics for 
$p(\vec e ,e ')X$
(energies of the initial and final electron and electron scattering angle) is 
contained in the single parameter $\epsilon$, for which we used three 
different 
values: $\epsilon=0,~1/2$ and 1. In order to extract the strong $k^2$ dependence 
of ${A}$, the "reduced" asymmetry ${A}_0=-{A}/|k^2|$ is shown.

In this picture one can see that the behavior of ${A}$  versus 
$k^2$ and $W$, in the region $1.08\le W\le 1.26$ GeV and in a wide region of 
momentum transfer $k^2$, is smooth everywhere and negative (note the $-1/|k^2|$ 
factor in the formula). Such a  behavior results from the isovector nature of 
the electroproduction processes which we have considered.

The role of the different contributions is illustrated in Figs. 7, 8 and 9.
In Fig. 7 (Fig. 8) the ratio of the cross sections $R_L^{(s)}$ and $R_T^{(s)}$ 
is 
reported as a function of W, for two fixed values of $|k^2|$, (a) $|k^2|=0.5$  
and (b)
$1.0$ (GeV/c)$^2$, for the reaction 
$\gamma^*+p\to p+\pi^0$ ($\gamma^*+p\to n+\pi^+$). 
The $\Delta$ contribution (dashed-dotted line) vanishes, while the Born terms 
(dotted line) give the largest contribution at forward angles. The contribution 
given by the vector meson ($\rho$ and $\omega$) exchange 
diagrams is not so essential here.

The different contributions to the total asymmetry A are shown in Fig. 9. Fig. 9 
is a projection of Fig. 6 showing the resulting reduced asymmetries 
${A}_0=-{A}/|k^2|$ as a function of $W$ at a fixed value of the virtual photon 
polarization  
$\epsilon=0.5$ and for two values of the momentum transfer (a) $|k^2|=0.5$ 
(GeV/c)$^2$; (b) $1.0$ (GeV/c)$^2$. The $\Delta$-contribution only is constant 
as a function of $W$ due to its isovector dominance, the vector-meson exchange 
gives a rather small contribution at low $W$ (below 1.2 $GeV$) and it is 
negligible above.
The full calculation gives values of A varying smoothly from $-7\cdot 10^{-5}$ 
at $W$=1.1 GeV (close to the elastic region) to   $-8\cdot 10^{-5}$ at $W$=1.25 
GeV, in the region of the $\Delta$ at  $|k^2|=1$ (GeV/c)$^2$.

One purpose of the present paper is to give an estimate of a possible 
contamination of the elastic peak by $\pi$-production and to assess its effect 
on the measured asymmetry. Although a detailed comparison of elastic and 
inelastic channels can only be made for specific geometries at the same incident 
energies and scattering angles (or equivalently, same $k^2$ and $\epsilon$), a 
rough estimate can be made using inclusive cross-sections.

Weak asymmetries have been calculated in the Standard Weinberg-Salam model 
assuming no strangeness in the nucleon. The results depend on the 
electromagnetic form factors for protons and neutrons, and give therefore 
different predictions 
depending on which values are taken. Calculations of ref.\cite{Na91} give 
$ A/|k^2|=-(1.4\div 1.5\cdot 10^{-5})$, at $|k^2|$=0.1 (GeV/c)$^2$ up to 
$-(2.2-2.8\cdot 10{-5})$, at $|k^2|$=0.3 (GeV/c)$^2$ for $E_e$= 2 GeV.

The SAMPLE Collaboration \cite{Sp00} has measured $A=-4.92\pm 0.61\pm 0.73 \cdot 
10^{-6}$ at $|k^2|$=0.1 (GeV/c)$^2$ and backward angle. Weak asymmetry 
calculations have been done  for the $G^0$ \cite{G0}, PVA4-Mainz \cite{PVA4} and 
HAPPEX \cite{An99} experimental conditions, predicting values ranging from 
$-0.3\cdot 10^{-5}$ at $|k^2|$=0.1 (GeV/c)$^2$ to $-2.0\cdot 10^{-5}$  at  
$|k^2|$=0.5 (GeV/c)$^2$ \cite{Be89}.

As one can see all calculations agree to predict negative asymmetries except one 
\cite{CG78} where small positive values have been obtained in an extension of 
the SM when taking right-handed doublets of $t$ and $d$ quarks into account (in 
this early paper, $sin^2(\theta_W)$ is taken equal to 1/3). Note that one can 
also find in this paper a simple explanation of the negative sign of the 
asymmetry in terms of $Z^0$ coupling to quarks and leptons.

Now comparing elastic scattering and inclusive $\pi$-production (Fig. 9), we see 
that they are both negative and of the same order of magnitude. Moreover $A$ is 
smaller in the region $W$=1.1 GeV (close to elastic scattering) and larger in 
the $\Delta $ region. Therefore we can conclude that a small admixture of 
$\pi$-production events in the region of the elastic peak, is not going to 
produce a large uncertainty in the elastic PV asymmetry. Specific cases can be 
computed using Fig. 6 or from the corresponding numerical values available from 
the authors.

\section{Conclusions}

We have derived the dependence of the  P-odd asymmetry for inclusive 
scattering 
of 
longitudinally polarized electrons by unpolarized protons with production of 
neutral and positive pions $\:p(\vec e,e')X$, with $X=p+\pi^0$ or $X=n+\pi^+$.
Using the known isotopic properties of the electromagnetic current for the 
$\gamma^*+p\to p+\pi^0$ and $\gamma^*+p\to n+\pi^+$ processes and the vector 
part of the hadronic weak neutral current for the $Z^*+p\to  p+\pi^0$ and 
$Z^*+n\to n+\pi^+$  processes, we derived an original expression for the 
inclusive asymmetry ${A}$. Without approximations, it is possible to disentangle 
the main 
contribution to ${A}$, which depends only on $k^2$. The calculation of  
${A}$ is then reduced to the analysis of specific isoscalar 
contributions 
to the electromagnetic currents. Such contributions appear, then, as 
corrections 
to the main term, which can be calculated exactly.

We have calculated the amplitudes 
for 
$\gamma^*+p\to N+\pi$, taking into account three standard contributions: 
Born+vector meson exchange+ $\Delta$-excitation. All the necessary 
parameters: 
interaction constants and different electromagnetic form factors are taken 
from 
other sources. Small adjustements in this basic approach were done in order to 
obtain a good description of the differential cross 
section data for the $\gamma+p\to p+\pi^0$ and $\gamma+p\to n+\pi^+$ reactions.
The model gives the vector part of the hadronic weak neutral current which is 
the main 
contribution to P-odd effects in $e+N\to e+N+\pi$.

The reduced weak asymmetry ${A}_0$ varies very little as a function of the 
two 
basic kinematical variables, $k^2$ and $W$.  In our approach this appears 
naturally 
from the fact that the isoscalar content of the electromagnetic current for 
$\gamma+N\to N+\pi$ is very small in the considered kinematical region. It is of 
the same sign and size as the $ep$ elastic PV asymmetry, and will, therefore, 
not much contribute to the experimental uncertainty of the former.

A straightforward extension of the present model would open a way to use P-odd 
observables in elastic and inelastic 
electron-proton scattering for the study of the axial 
contributions. We 
plan to discuss this in a forthcoming paper.

\section{Appendix}

\subsection{Born contribution: s-channel}

The scalar amplitudes, for $\gamma^*+p\to p+\pi^0$ are defined as

$$f_{1s}=f_{3s}=-\displaystyle\frac{g}{W-m}\displaystyle\frac
{|\vec k||\vec q|}{(E_1+m)(E_2+m)}\left [ 
F_{1p}(k^2)+F_{2p}(k^2)\displaystyle\frac{W+m}{2m}\right ],$$
$$f_{2s}=\displaystyle\frac{g}{W-m}
\left [ F_{1p}(k^2)-F_{2p}(k^2)\displaystyle\frac{W-m}{2m}\right ],
$$
$$
+\frac{g}{W-m}\displaystyle\frac
{\vec k\cdot\vec q}{(E_1+m)(E_2+m)}\left [ 
F_{1p}(k^2)+F_{2p}(k^2)\displaystyle\frac{W+m}{2m}\right ],$$
$$f_{4s}=0,$$
$$f_{5s}=\displaystyle\frac{g}{(W+m)(E_1+m)}
\left [-F_{1p}(k^2)+F_{2p}(k^2)\displaystyle\frac{E_1+m}{2m}\right ],
$$
$$
f_{6s}=\displaystyle\frac{g}{(W-m)(E_2+m)}
\displaystyle\frac{|\vec k|}{|\vec q|}\left [ - 
F_{1p}(k^2)+F_{2p}(k^2)\displaystyle\frac{E_1-m}{2m}\right ],$$
with $ |\vec k|=\sqrt{E_1^2-m^2}$ and $|\vec q|=\sqrt{E_2^2-m^2}.$

\subsection{Born contribution: u-channel}
$$f_{1u}=\displaystyle\frac{g|\vec k||\vec q|}{u-m^2}
\left \{ F_{1p}(k^2)
\displaystyle\frac{W+m}{(E_1+m)(E_2+m)}\right .$$
$$\left .-\displaystyle\frac{F_{2p}(k^2)}{2m(E_1+m)}
\left [ W+m+\left (m+\displaystyle\frac
{m^2-k^2}{W}\right )\displaystyle\frac{E_{\pi}}{E_2+m}+
\displaystyle\frac{2\vec k\cdot\vec q}{E_2+m}\right ]\right \},$$
$$f_{2u}=\displaystyle\frac{g}{u-m^2}\left [F_{1p}(k^2)\left (W-m+
\vec k\cdot\vec q\displaystyle\frac{W+m}{(E_1+m)(E_2+m)}\right ) \right .$$
$$-\displaystyle\frac{F_{2p}(k^2)}{2m}\left ((E_1-m)(W-m)+\tilde k_0
\left( m+\displaystyle\frac{m^2-m_{\pi}^2}{W}\right )\right .$$
$$\left .\left .\left (-2\vec k\cdot\vec q +
\left (m+\displaystyle\frac{m^2-k^2}{W}\right )\left (
m+\displaystyle\frac{m^2-m_{\pi}^2}{W}\right )\right )
\displaystyle\frac{\vec k\cdot\vec q}{(E_1+m)(E_2+m)}\right )\right ],$$
$$f_{3u}=\displaystyle\frac{g}{u-m^2}\displaystyle\frac{g|\vec k||\vec q|}
{E_1+m}
\left [ -\displaystyle\frac{F_{1p}(k^2)}{E_2+m}
\left (
-m+\displaystyle\frac{m_{\pi}^2-m^2}{W}\right )\right .$$
$$-\displaystyle\frac{F_{2p}(k^2)}{2m}\left .
\left (-W-m+\left ( m+\displaystyle\frac
{m^2-k^2}{W}\right )\displaystyle\frac{E_{\pi}}{E_2+m}
+\displaystyle\frac{2\vec k\cdot\vec q}{E_2+m}\right )\right ],$$
$$f_{4u}=\displaystyle\frac{g}{u-m^2}(E_2-m)\left [-2F_{1p}(k^2)+F_{2p}(k^2)
\left (-1+\displaystyle\frac{W}{m}\right)\right ],$$
$$f_{5u}=-\displaystyle\frac{g}{(u-m^2)(E_1+m)}\left [ \left ( -F_{1p}(k^2)
+F_{2p}(k^2)\displaystyle\frac{E_1+m}{2m}\right )
\left (m+ \displaystyle\frac{m_{\pi}^2 -m^2}{W}\right ) \right .$$
$$ \left .+F_{2p}(k^2)\displaystyle\frac{\vec k\cdot\vec q}{m} \right ],$$
$$f_{6u}=\displaystyle\frac{g}{(u-m^2)(E_2+m)}
\displaystyle\frac{|\vec q|}{|\vec k|}
\left [-\left (F_{1p}(k^2)+F_{2p}(k^2)\displaystyle\frac{E_1-m}{2m}
\right )\left( m- \displaystyle\frac{m_{\pi}^2 -m^2}{W}\right )+
F_{2p}(k^2)\displaystyle\frac{\vec k\cdot\vec q}{m}\right],$$
where
$$u-m^2=k^2-2\tilde k_0E_2-2\vec k\cdot\vec q,~~
\tilde k_0=\displaystyle\frac{W^2+k^2-m^2}{2W},~\mbox{and} ~E_{\pi}=W-E_2.$$
\subsection{Vector meson exchange: t-channel}
$$f_{1V}=g_{V\pi\gamma^*}(k^2)g_V
\displaystyle\frac{|\vec k||\vec q|}{m_V(t-m_V^2)}$$
$$\left \{ \left [1+\left (1+\displaystyle\frac{W}{m} \right)\kappa_V
\right ]\left (-1+\displaystyle\frac{\vec k\cdot\vec q}{(E_1+m)(E_2+m)}
\right)
+(1+\kappa_V)(W+m)\left (\displaystyle\frac{1}{E_1+m}+
\displaystyle\frac{1}{E_2+m}\right )\right \},$$
$$f_{2V}=g_{V\pi\gamma^*}(k^2)\displaystyle\frac{g_V}{m_V(t-m_V^2)}$$
$$\left \{ (1+\kappa_v)
\left [ \tilde k_0(E_2-m)+
E_{\pi}(E_1-m)-\vec k\cdot\vec q
\left(
\displaystyle\frac{\tilde k_0}{E_1+m}+
\displaystyle\frac{E_{\pi}}{E_2+m}
\right ) \right ]\right.$$
$$\left .  +
\left [1+\left(1+\displaystyle\frac{W}{m}\right )\kappa_V\right ]
\displaystyle\frac{\vec k^2\vec q^2-(\vec k\cdot\vec q)^2}{(E_1+m)(E_2+m)} 
\right \},$$
$$f_{3V}=g_{V\pi\gamma^*}(k^2)\displaystyle\frac{g_V|\vec k||\vec 
q|}{m_V(t-m_V^2)}$$
$$ \left \{(1+ \kappa_V)\displaystyle\frac{E_{\pi}}{E_2+m}+
\left [1+\left(1+\displaystyle\frac{W}{m}\right )\kappa_V\right ]
\displaystyle\frac{(\vec k\cdot\vec q)}{(E_1+m)(E_2+m)}
\right \},$$
$$f_{4V}=-g_{V\pi\gamma^*}(k^2)\displaystyle\frac{g_V}{m_V}$$
$$\displaystyle\frac{(E_1+m)(E_2+m)}{t-m_V^2}
\left [1+\left(1+\displaystyle\frac{W}{m}\right )\kappa_V +
\displaystyle\frac{(1+\kappa_V)}{E_1-m}\tilde k_0\right ],$$
$$
f_{5V}=g_{V\pi\gamma^*}(k^2)\displaystyle\frac{g_V}{m_V(t-m_V^2)}
\displaystyle\frac{1+\kappa_V}{E_1+m}
\left [ t+\left ( k^2-m^2_{\pi}\right )
\displaystyle\frac{m}{W}\right ],
$$
$$f_{6V}=-g_{V\pi\gamma^*}(k^2)\displaystyle\frac{g_V}{m_V(t-m_V^2)}
\displaystyle\frac{1+\kappa_V}{E_1+m}
\displaystyle\frac{|\vec k||\vec q|}{m_V(t-m_V^2)}
\left [t+\left ( k^2-m^2_{\pi}\right )\displaystyle\frac{m}{W}\right ],
$$
where $$t-m_V^2=m_{\pi}^2-m_V^2-2\tilde k_0E_{\pi}+2\vec k\cdot\vec q+k^2.$$
\subsection{One pion contribution: t-channel}
$$f_{1t}=f_{2t}=0,$$
$$f_{3t}=g\displaystyle\frac{2|\vec k||\vec q|}{t-m_{\pi}^2}
\displaystyle\frac{F_{\pi}(k^2)}{E_1+m},$$
$$f_{4t}=-2g\displaystyle\frac{E_2-m}{t-m_{\pi}^2}
F_{\pi}(k^2),$$
$$
f_{5t}=-\displaystyle\frac{g}{t-m_{\pi}^2}
F_{\pi}(k^2)\displaystyle\frac{2E_{\pi}-k_0}{E_1+m},
$$
$$f_{6t}=-\displaystyle\frac{g}{t-m_{\pi}^2}
F_{\pi}(k^2)
\displaystyle\frac{|\vec k||\vec q|}{t-m_{\pi}^2}
\displaystyle\frac{2E_{\pi}-k_0}{E_2+m},$$
where $$F_{\pi}(k^2)=\displaystyle\frac{1}{1-k^2/m^2_{\rho} }$$
is the pion electromagnetic form factor, given in the framework of VDM, 
$m_{\rho}=0.77$ GeV is the $\rho$-meson mass.

\subsection{Calculation of the isoscalar amplitudes $f_i^{(s)}(\gamma^*p\to 
p\pi^0)$}
The isoscalar amplitudes are:
$$f_i^{(s)}(\gamma^*p\to p\pi^0)=-f_{i,s}^{(s)}-f_{i,u}^{(s)}-f_{i,\rho}$$
where the contributions $f_{i,s}^{(s)}$ and $f_{i,u}^{(s)}$
are determined by the corresponding formulas, 
with the following substitutions:
$$F_{1p}\to F_{1s}=\displaystyle\frac{F_{1p}+F_{1n}}{2},$$
$$F_{2p}\to F_{2s}=\displaystyle\frac{F_{2p}+F_{2n}}{2},$$
with 
$$F_{1n}=\displaystyle\frac{G_{En}-\tau G_{Mn}}{1-\tau},~~
F_{2n}=\displaystyle\frac{-G_{En}+G_{Mn}}{1-\tau}.$$

\subsection{Gauge invariance of the suggested model}

In the framework of the considered model, for the process of neutral pion 
electroproduction, $e+p \to e+p+\pi^0$, the corresponding hadronic 
electromagnetic current is conserved:$k\cdot {\cal J}^{(em)}_\mu=0$
for any form factor in $\gamma^*NN$,$\gamma^*\pi\pi$,$\gamma^*V\pi$,
and $\gamma^*N\Delta$-vertices.

In case of charged pion electroproduction, $e+p \to e+n+\pi^+$, a special 
contribution must be added to the matrix element:
$$\Delta{\cal M}=-\sqrt{2}g\displaystyle\frac{e\cdot 
k}{k^2}\gamma_5(F_{1p}-F_{1n}-F_{\pi})$$
which results in additional contributions to the scalar amplitudes: 
$\Delta f_i(\gamma p\to n\pi^+)$:
$$\Delta f_1=\Delta f_2=\Delta f_2=\Delta f_4=0
$$
$$\Delta f_5=\sqrt{2}g(E_1-m)[F_{1p}(k^2)-F_{1n}(k^2)-F_{\pi}(k^2)]/k^2$$
$$\Delta f_6=-\sqrt{2}g(E_2-m)[F_{1p}(k^2)-F_{1n}(k^2)-F_{\pi}(k^2)]/k^2$$


\begin{table*}
\begin{tabular}{|c|c|c|c|c|c|c|}
Reaction &$a_s$ &$a_u$&$a_t$&$a_\rho$&$a_\omega$&$a_\Delta$\\
\hline\hline
$\gamma^*+p\to p+\pi^0$&-1&-1&0&+1&+1&$\sqrt{2}$\\
$\gamma^*+p\to n+\pi^+$&$\sqrt{2}$&$\sqrt{2}$&$\sqrt{2}$&$\sqrt{2}$&0&+1\\
\end{tabular}
\caption{Numerical coefficients for the different contributions to the Feynmann 
diagrams  }

\label{tab1}
\end{table*}


\begin{figure}
\mbox{\epsfxsize=15.cm\leavevmode\epsffile{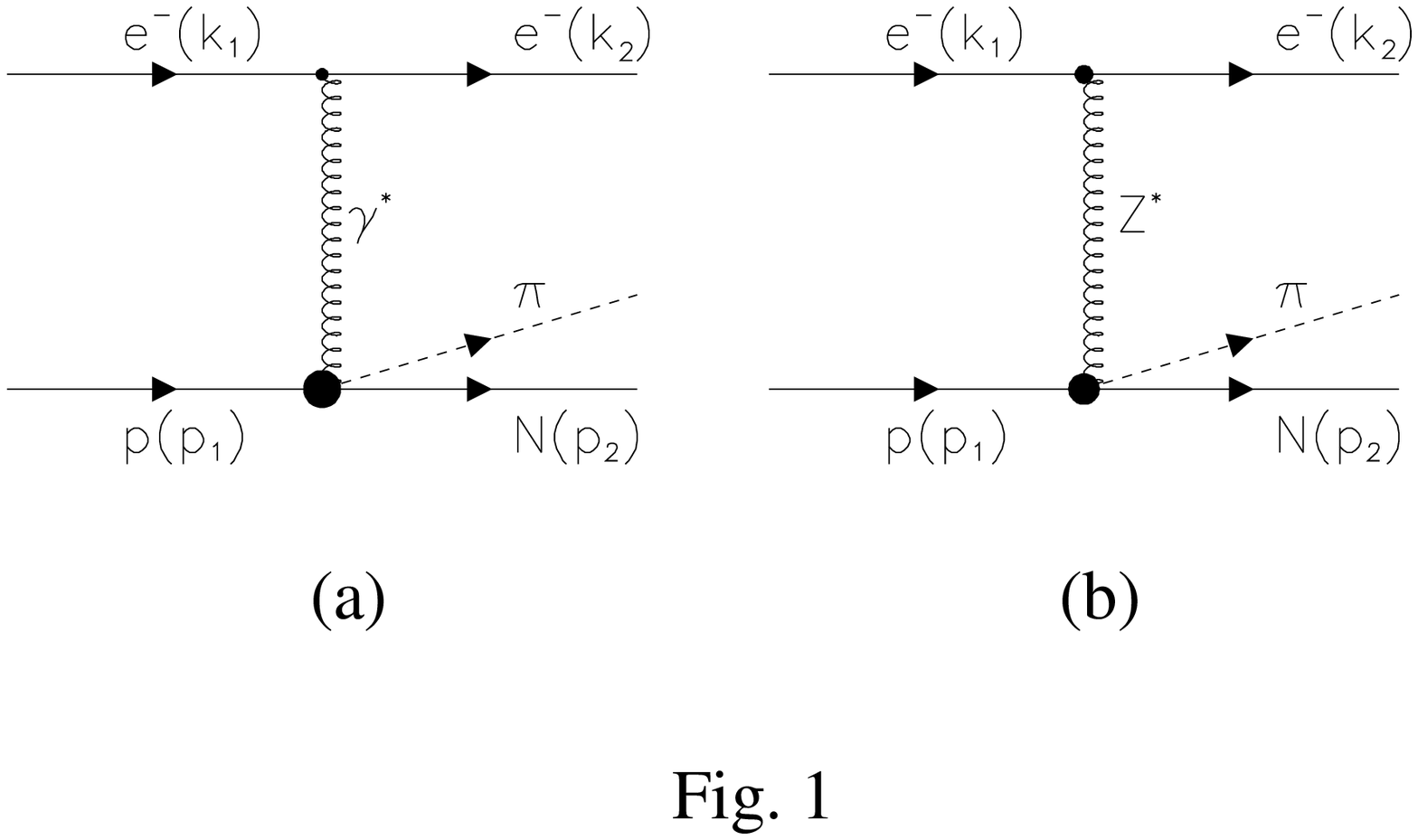}}
\caption{ Feynman diagrams for $\gamma ^*$- and $Z^*$-boson exchanges in the 
processes $e^-+p\rightarrow e^-+N+\pi$.}
\end{figure}
\begin{figure}
\mbox{\epsfxsize=15.cm\leavevmode\epsffile{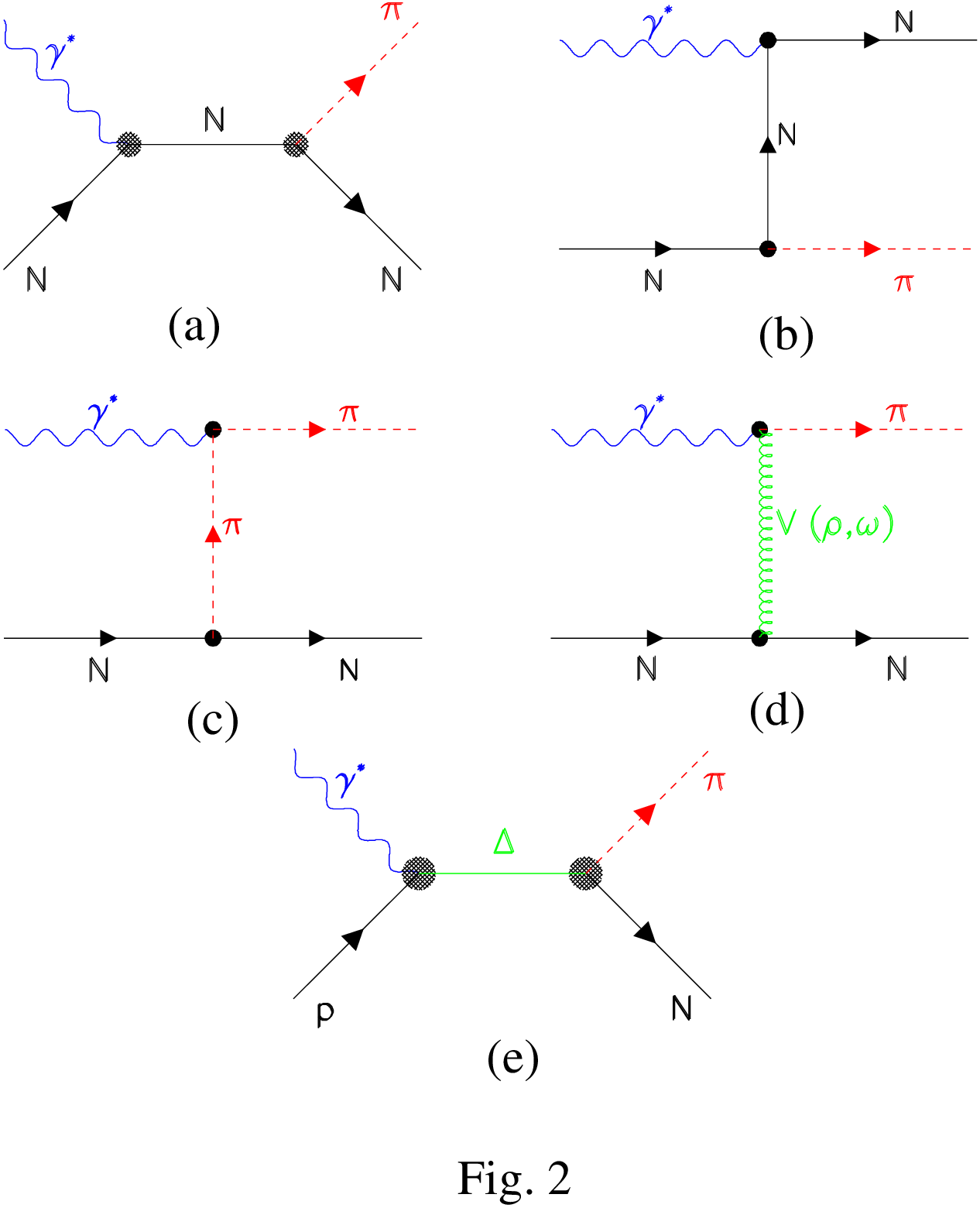}}
\caption{ Feynman diagrams for $\gamma^*+p\rightarrow N+\pi$- 
processes.}
\end{figure}
\begin{figure}
\mbox{\epsfxsize=15.cm\leavevmode\epsffile{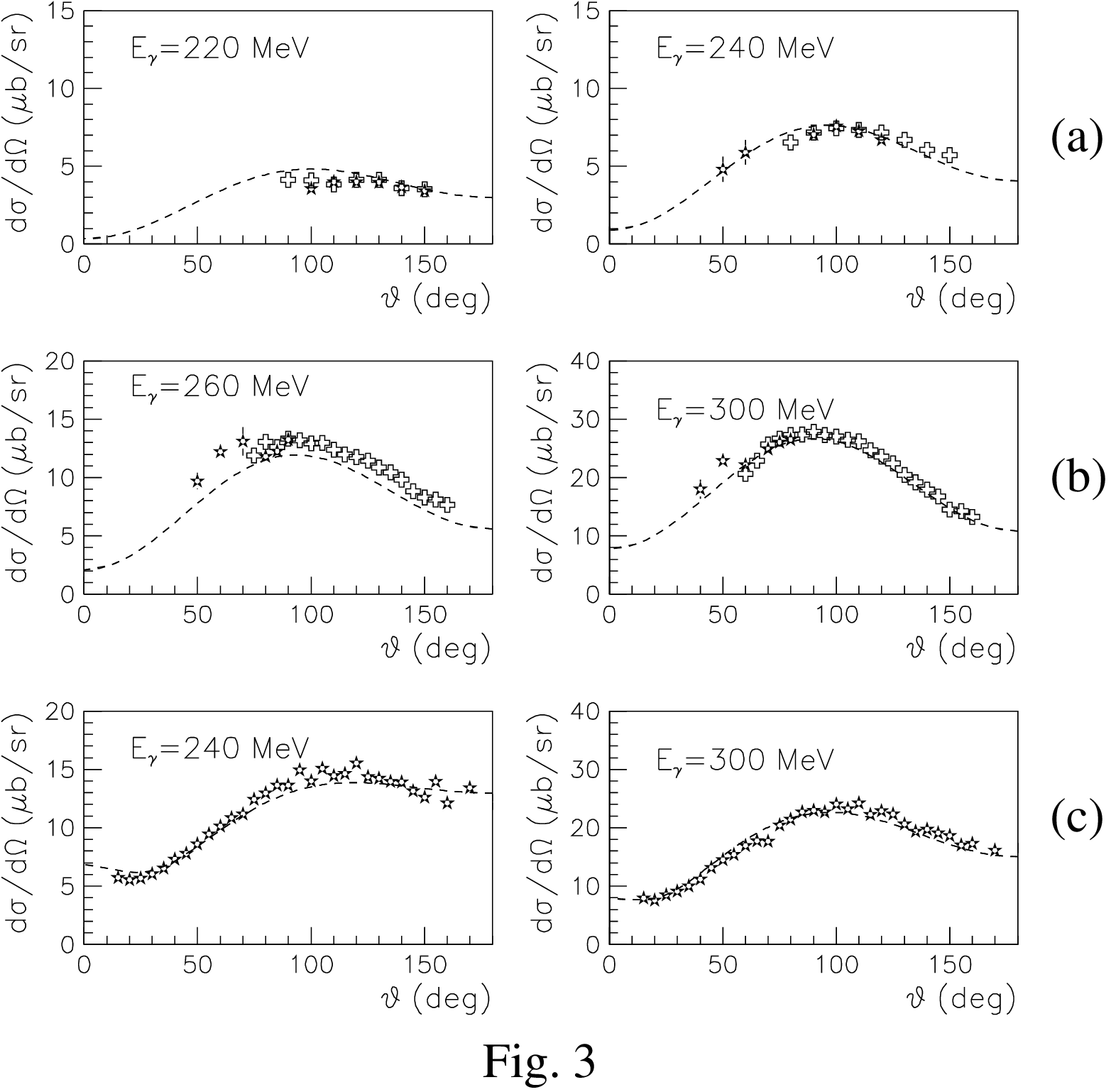}}
\caption{The angular dependence of the differential cross sections for the 
photoproduction processes: (a) and (b) $\gamma^*+p\rightarrow p+\pi^0$; open 
stars: data from ref.\protect\cite{FH72}, open crosses: data from 
ref.\protect\cite {GH74},
(c) $\gamma^*+p\rightarrow n+\pi^+$; data are from ref.\protect\cite{FF71}; the 
dashed line is the prediction of the 
present model.}
\end{figure}
\begin{figure}
\mbox{\epsfxsize=15.cm\leavevmode\epsffile{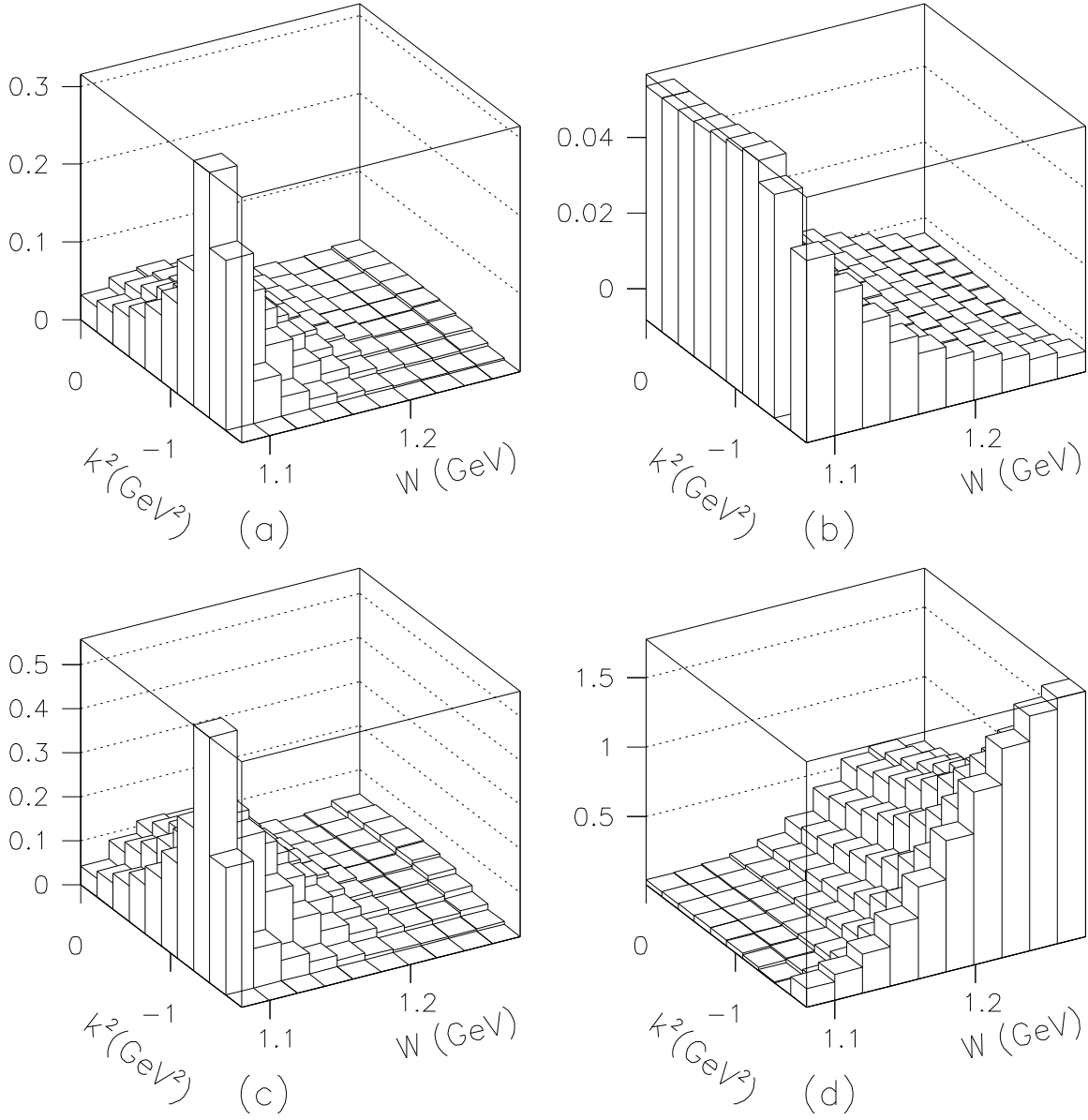}}
\caption{The $k^2$ and $W$-dependences of the ratios of the total cross 
sections 
for the $e^-+p\to e^-+p+\pi^0$ reaction: 
(a) $R_L^{(s)}(k^2,W)=\sigma_L^{(s)}(k^2,W)/\sigma_T(k^2,W)$; 
(b) $R_T^{(s)}(k^2,W)=\sigma_T^{(s)}(k^2,W)/\sigma_T(k^2,W)$; 
(c) $R_{LT}(k^2,W)=\sigma_L(k^2,W)/\sigma_T(k^2,W)$; 
(d) $R_{np}=\sigma_T(p\pi^0)/\sigma_T(n\pi^+)$. 
}
\end{figure}
\begin{figure}
\mbox{\epsfxsize=15.cm\leavevmode\epsffile{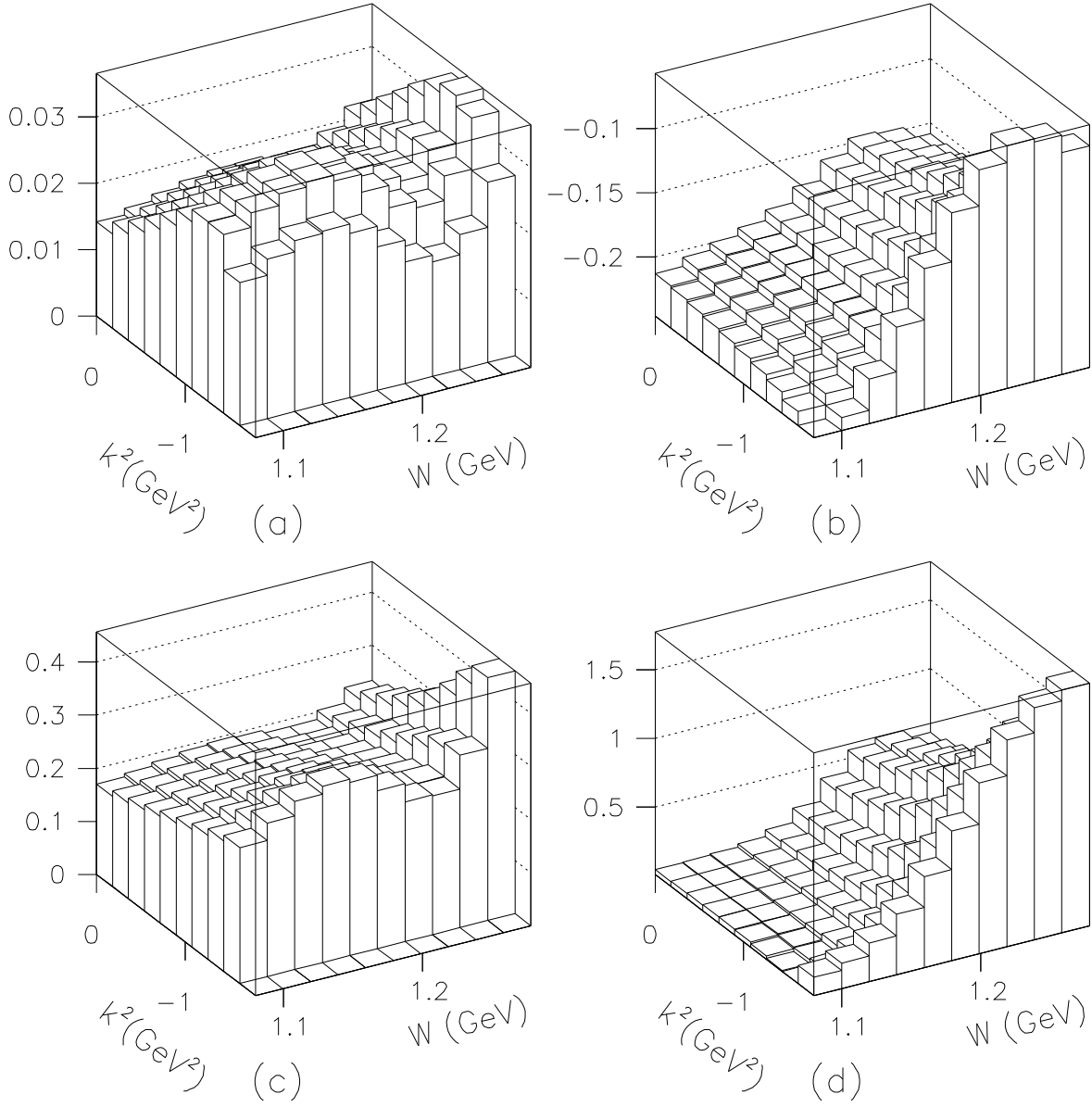}}
\caption{The $k^2$ and $W$-dependences of the ratios of the total cross 
sections 
for the $e^-+p\to e^-+n+\pi^+$ reaction. Same conventions as in Fig. 4.
}
\end{figure}
\begin{figure}
\mbox{\epsfxsize=15.cm\leavevmode\epsffile{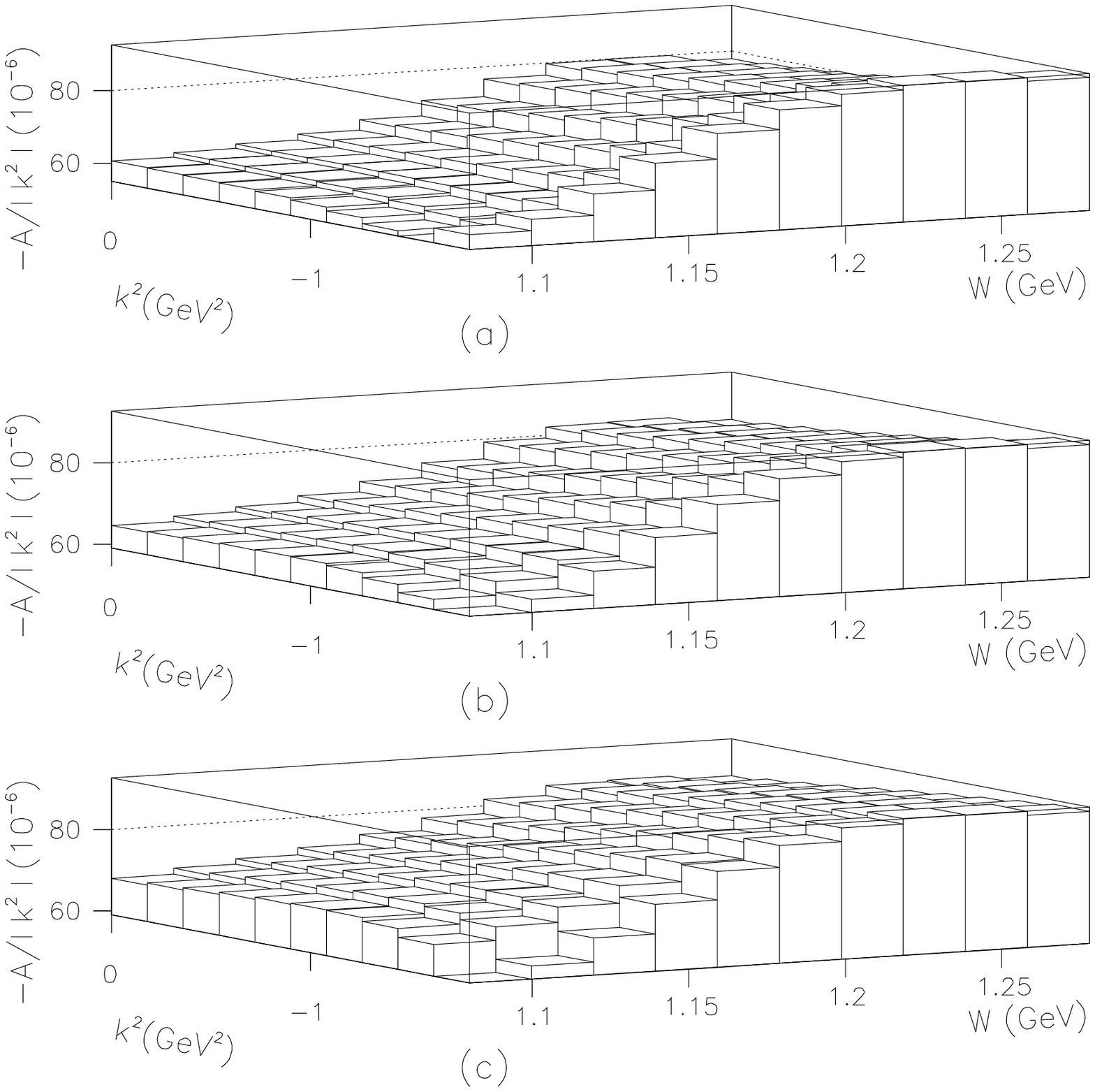}}
\caption{The $k^2$ and $W$ dependences of the reduced asymmetry 
${A}_0=-{A}/|k^2|$ (where A is the theoretical asymmetry 
to be compared to experimental data)
for 
 $p(\vec e,e')X$ at three different values of the virtual photon polarization  
$\epsilon$: (a) $\epsilon=0$; (b) $\epsilon=0.5$; (c) $\epsilon=1$.
}
\end{figure}
\begin{figure}
\mbox{\epsfxsize=15.cm\leavevmode\epsffile{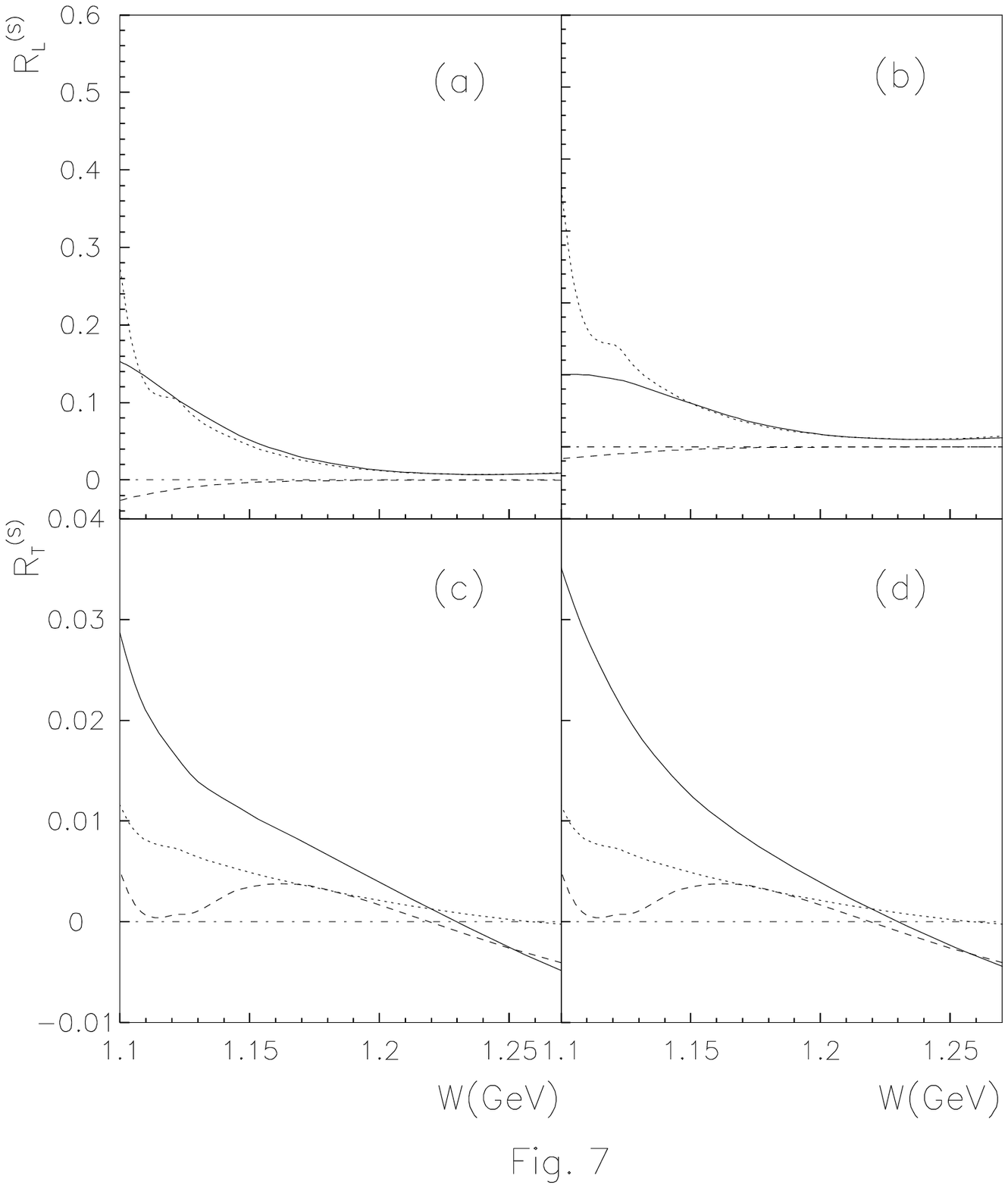}}
\caption{The  $W$-dependence of the ratios $R_L^{(s)}(k^2,W)$ and 
$R_T^{(s)}(k^2,W)$ for fixed values of $k^2$ a)) and c) $-k^2=0.5$ (GeV/c)$^2$; 
b) and d) $-k^2=1.0$ (GeV/c)$^2$ 
for the $e^-+p\to e^-+p+\pi^0$ reaction. The curves represent the full 
calculation (full line), $\Delta$-contribution only (dashed-dotted line), 
$\Delta$ + Born terms (dashed line), $\Delta$ + vector mesons (dotted line).
}
\end{figure}
\begin{figure}
\mbox{\epsfxsize=15.cm\leavevmode\epsffile{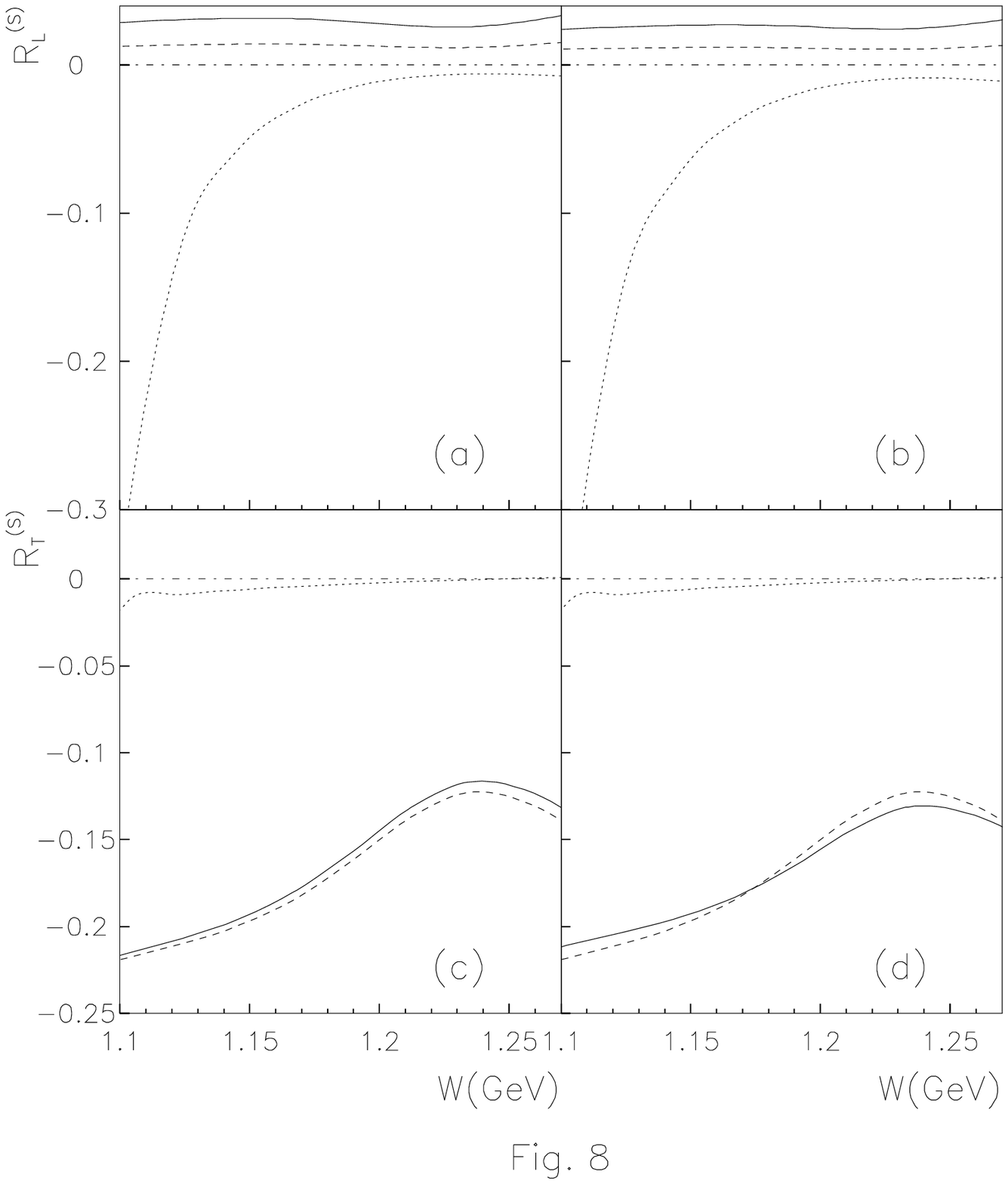}}
\caption{The same as Fig. 7, for the $e^-+p\to e^-+n+\pi^+$ reaction
}
\end{figure}
\begin{figure}
\mbox{\epsfxsize=15.cm\leavevmode\epsffile{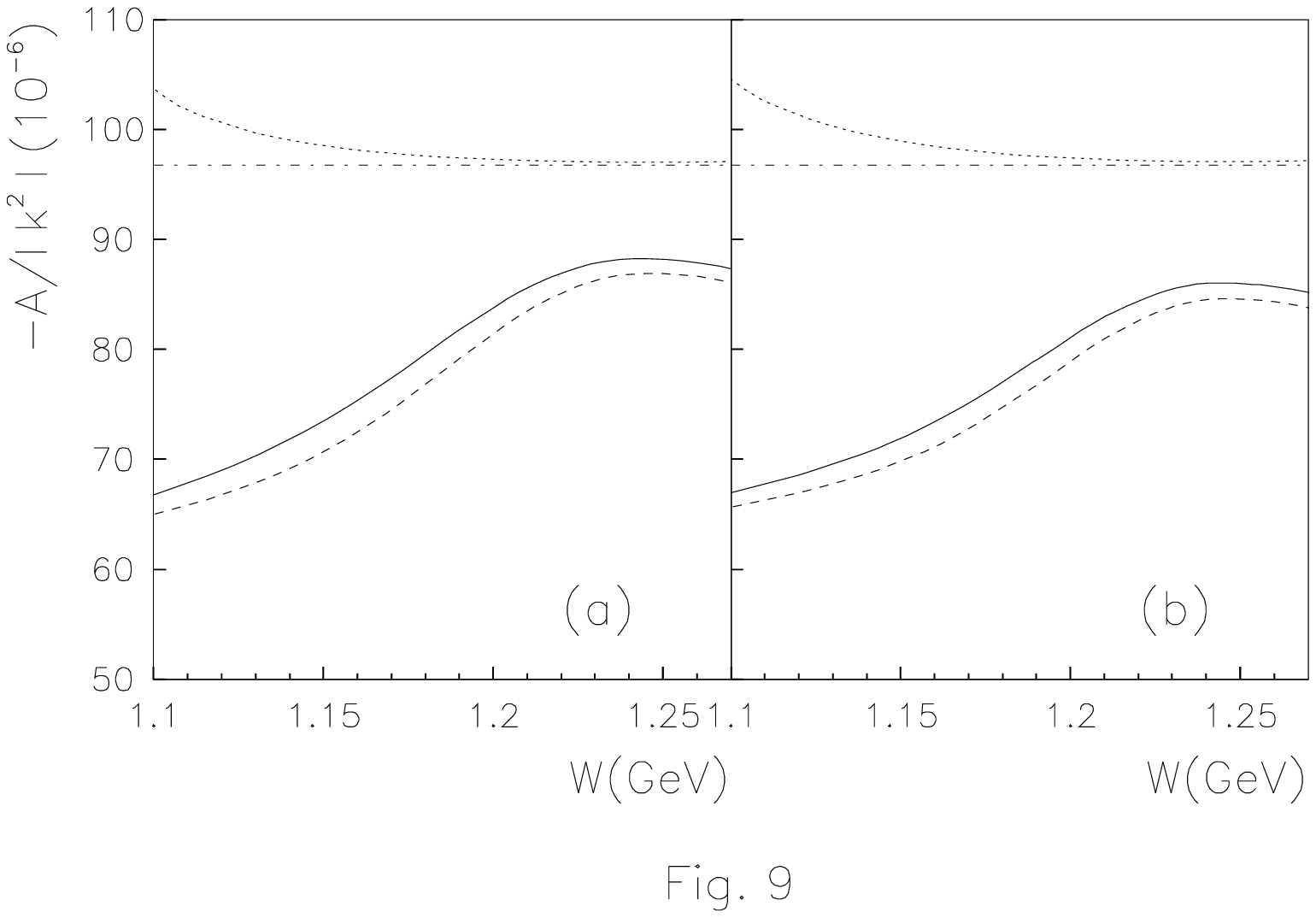}}
\caption{The  $W$-dependence of the reduced asymmetry ${A_0}$ for 
$\epsilon=0.5$ and two values of $k^2$: (a) $-k^2=0.5$; (b) $1.0$ (GeV/c)$^2$. 
Same conventions  as in Fig. 7.
}
\end{figure}

\end{document}